\documentclass[prx, reprint, notitlepage,
nofootinbib,
superscriptaddress,
floatfix]{revtex4-1}
\usepackage{gpkmac}

\usepackage{xcolor}

\usepackage{txfonts}
\usepackage{bbold}
\usepackage[normalem]{ulem}

\def\be{\begin{equation}}
\def\ee{\end{equation}}
\newcommand{\st}[1]{}

\begin{document}

%---------------------------------------------------------

% \preprint{arXiv:xxxx.xxxxx[cont-mat]}

\title{Cross Entropy Benchmark for Measurement-Induced Phase Transitions}

\author{Yaodong Li}
% \email{liyd@stanford.edu}
\affiliation{Department of Physics, University of California, Santa Barbara, CA 93106}
\affiliation{Department of Physics, Stanford University, Stanford, CA 94305}

\author{Yijian Zou}
% \email{zouy@stanford.edu}
\affiliation{Department of Physics, Stanford University, Stanford, CA 94305}

\author{Paolo Glorioso}
% \email{paolog@stanford.edu}
\affiliation{Department of Physics, Stanford University, Stanford, CA 94305}

\author{Ehud Altman}
% \email{ehud.altman@berkeley.edu}
\affiliation{Department of Physics, University of California, Berkeley, CA 94720}

\author{Matthew P. A. Fisher}
% \email{mpaf@kitp.ucsb.edu}
\affiliation{Department of Physics, University of California, Santa Barbara, CA 93106}

% \date{January 3, 2023}
% \date{March 22, 2023}
\date{May 6, 2023}

\begin{abstract}

We investigate prospects of employing the linear cross entropy to experimentally access   measurement-induced phase transitions (MIPT) without requiring any postselection of quantum trajectories.
For two random circuits that are identical in the bulk but with different initial states, 
the linear cross entropy $\chi$ between the bulk measurement outcome distributions in the two circuits acts as an order parameter, and can be used to distinguish the volume law from area law phases. 
In the volume law phase (and in the thermodynamic limit) the bulk measurements cannot distinguish between the two different initial states, and $\chi=1$.  In the area law phase $\chi <1$.
For circuits with Clifford gates, we provide numerical evidence that $\chi$ can be sampled to accuracy $\varepsilon$ from 
$O(1/\varepsilon^2)$
trajectories, by running the first circuit on a quantum simulator  \emph{without postselection}, aided by a classical simulation of the second.
We also find that for weak depolarizing noise
{the signature of} the MIPT is still present for intermediate system sizes.
In our protocol we have the freedom of choosing initial states
such that the ``classical'' side can be simulated efficiently, while simulating the ``quantum'' side is still classically hard. 

\end{abstract}

\maketitle

%-----------------------------------------------------

% {\hypersetup{linktocpage} \tableofcontents}

% \YL{
% TODO:

% Matteo pointed out a minor mistake in Appendix B

% Yijian's Haar numerics

% Michael Gullans suggestion

% my simplified protocol calculation
% }

% \tableofcontents

\emph{Introduction.} ---
Open quantum dynamics can host a rich phenomenology, including a family of measurement-induced phase transitions (MIPT) in the scaling of entanglement along quantum trajectories in monitored systems~\cite{nahum2018hybrid, nandkishore2018hybrid, li2018hybrid, choi2019qec, gullans2019purification, andreas2019hybrid, choi2019spin}.
The MIPT {is a basic phenomenon in many-body quantum dynamics and} occurs generically in a number of different models~\cite{cao2018monitoring, li2019hybrid, szyniszewski2019measurement,Tang2019, nahum2019majorana,  vasseur2020mft, barkeshli2020symmetric, sang2020protected,  ippoliti2020measurementonly, chenxiao2020nonunitary, ashida2020continuous,diehl2020trajectory,  pal_lunt_2020_mbl_hybrid, sagar2020volume, dalmonte2020twoplusonedim,  nahum2020alltoall,   bao2021enriched, vasseur2021chargesharpening, barratt2021sharpening}, yet its experimental observation can be challenging even on an error-corrected quantum computer, due to the so-called ``postselection problem''.
Quantum trajectories are labeled by the measurement history $\bs{m}$, whose length is extensive in the space-time volume $V$ of the circuit; thus, the number of possible trajectories $\bs{m}$ is exponential in $V$, but they each occur with roughly the same probability.
On the other hand, one needs multiple copies of the same $\bs{m}$ in order to verify any quantum entanglement; and then many different $\bs{m}$ to perform a proper statistical average.
On a quantum simulator there is no general recipe for producing such copies other than running the quantum circuit many times and waiting until the measurement results coincide (``postselection'').
% In other words, the prepration of the output state is not readily ``repeatable'', and naively
Naively,
$O(e^V)$ runs of the circuit are required, % to generate multiple copies,
thus severely restricting the scalability of such experiments.
Nevertheless, in an impressive recent experiment that carries out postselection~\cite{koh2022experimental}, the MIPT is observed on small scale superconducting quantum processors.

The exponential postselection overhead has previously been shown to be avoidable in two cases.
First, when only Clifford circuits are considered, the entanglement can be verified by ``decoding'' the circuit, either through a full classical simulation within the stabilizer formalism~\cite{monroe2021TrappedIonCliffordTransition} or via machine learning~\cite{dehghani2022neuralnetwork}.
With machine learning the authors claim that  ``decoding" is possbile also beyond Clifford circuits, although this has yet to be explored in detail.
Second, when the non-unitary (monitored) dynamics is a spacetime dual of a unitary one~\cite{ippoliti2020postselectionfree, ippoliti2021fractal, grover2021rotation}, postselection is partially ameliorated, and
correspondences between unitary dynamics and monitored dynamics can be made.
%\footnote{We bring the reader's attention to a recent proposal of revealing the MIPT using ``pre-selection''~\cite{diehl2022preselection}.}

% \EA{** Is the word "another" in the first sentence below necessary? I think the dual unitary networks do not have an MIPT and in any case are not resource efficient. I also doubt that the machine learning protocol is efficient in the general case ***  }

\begin{figure}[t]
    \centering
    \includegraphics[width=.5\textwidth]{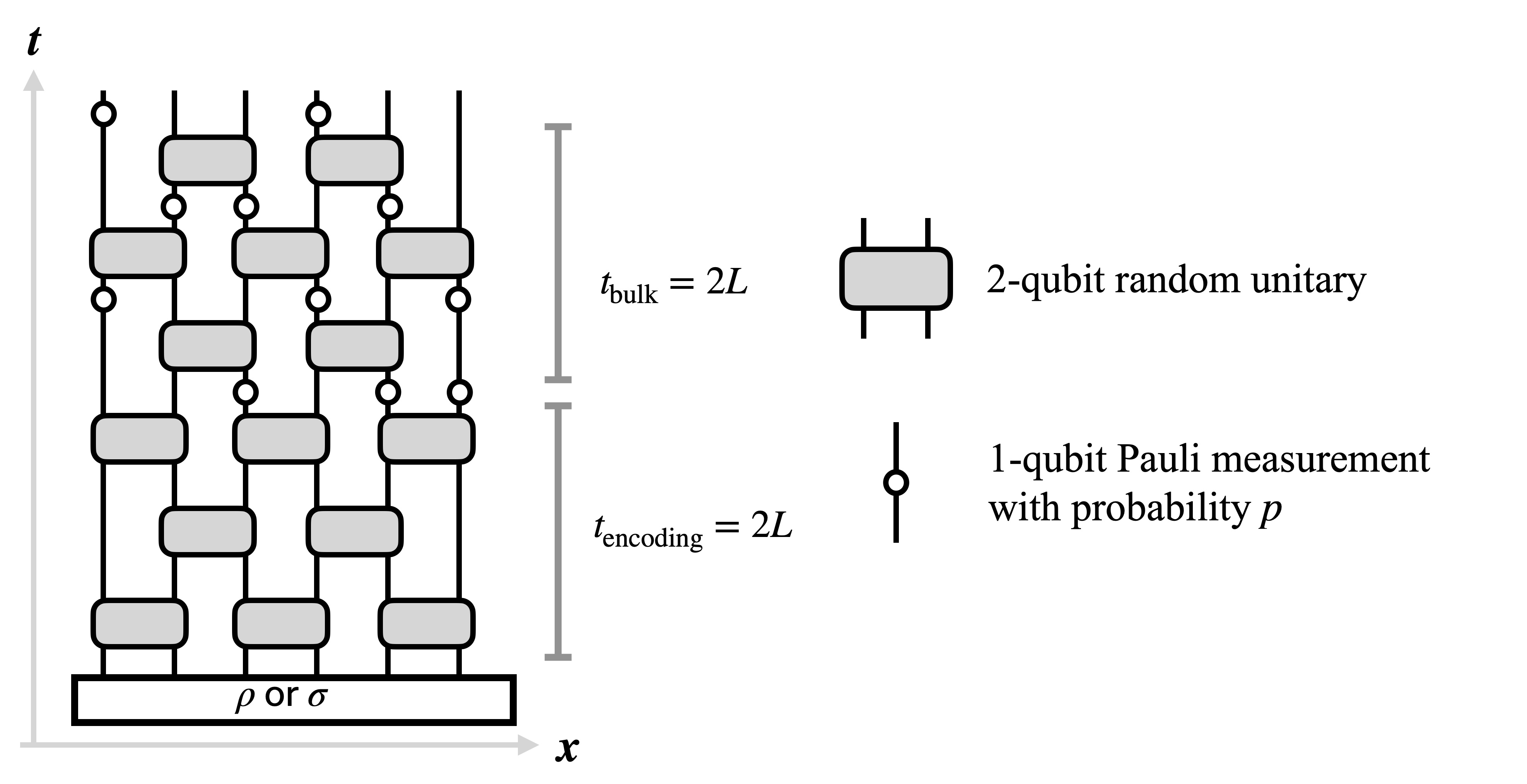}
    \caption{The layout of the hybrid circuit considered in this paper.
    Different from the usual setup~\cite{li2019hybrid}, we have an additional ``encoding'' stage before the hybrid evolution for time $t_{\rm encoding} = 2L$, following Ref.~\cite{gullans2019purification}.
    We call the evolution after the encoding stage the ``circuit bulk'', which lasts for another $t_{\rm bulk} = 2L$.
    The total circuit time is $T = t_{\rm encoding} + t_{\rm bulk} = 4L$.
    We will compare two different initial states $\rho$ and $\sigma$ (left unspecified for the moment) undergoing the same circuit evolution.
    }
    \label{fig:circuit}
\end{figure}

Here we propose a resource efficient experimental protocol for verifying the MIPT in random circuits, by estimating
% an order parameter known as
the ``linear cross entropy'' (denoted $\chi$) between
% measurement histories
the probability distribution of (bulk circuit) measurement outcomes $\bs{m}$
in two circuits with the same bulk but different initial states, $\rho$ and $\sigma$. 
% \st{This quantity was previously}
% \st{Similar quantities have been}
Closely related quantities have been discussed previously~\footnote{
In particular, Ref.~\cite{choi2019spin} proposed the Fisher information, quantifying the change in the bulk measurement outcome distribution when the initial state is slightly perturbed.
Ref.~\cite{gullans2019scalable} proposed the entropy of a reference qubit as a boundary order parameter, where the reference qubit is initially maximally-entangled with the system and gets purified under measurements.
Both quantities are are akin to a boundary magnetization, although in a different stat mech model, as we discuss in the Supplemental Material~\cite{SM}.
Ref.~\cite{gullans2019scalable} also considered the purification a reference qubit after an encoding stage is applied, in a way similar to Fig.~\ref{fig:circuit}.}.
% In terms of the spin model description of the MIPT~\cite{choi2019spin, andreas2019hybrid}, $\chi$ can be understood as a boundary correlation function. \YZ{I wonder whether this sentence should really be here. It is not conceptually necessary in the introduction part and not technically correct (what we really have in appendix is $\bar{\chi}$.) }
In particular, as we establish both numerically and analytically, in the thermodynamic limit the linear cross entropy (when suitably normalized) is $1$ in the volume law phase, and equals a nonuniversal constant smaller than $1$ in the area law phase.
Thus, the MIPT can also be viewed as a phase transition in the distinguishability of two initial states, when the bulk measurement outcomes are given.
{In particular, the two initial states become essentially indistinguishable when measurements are below a critical density.}
% Indeed, the MIPT is a transition in the bulk structure of the quantum circuit, where the scaling of the entanglement entropy in the output state is just one of its many ramifications.
% A full appreciation of this fact may lead to other ways of probing the transition experimentally.

The definition of $\chi$ includes contributions from
all samples of $\bs{m}$, and to estimate $\chi$ no postselection is involved.
However, as we discuss below,
estimating $\chi$ usually requires an exponentially long classical simulation, thus not scalable.
% In Sec.~\ref{sec:numerical_results_Clifford},
Below, we show that when the classical simulation becomes scalable in Clifford circuits,
$\chi$ can be efficiently sampled by running the $\rho$-circuit on a quantum simulator, aided by a classical simulation of the $\sigma$-circuit.
% In particular,
%\st{Within this protocol,}
% For a fixed circuit we estimate that the number of samples of $\bs{m}$ scales as
% % $\mathrm{poly}(1/\varepsilon)$.
% $1/\varepsilon^2$, where $\varepsilon$ is the error of the estimation of $\chi$.
We provide numerical evidence that $\chi$ is an order parameter for the MIPT (i.e. $\chi = 1$ in the volume law phase and $\chi <1$ in the area law phase).
% , and simulate the effect of depolarizing noise.
% We find that depolarizing noise generically decreases $\chi$ regardless of $p$, and in the thermodynamic limit we expect no transition for any finite noise rate.
% \YZ{Do we establish or just expect that there is no transition with any noise?}

% Notably, our protocol does not require any postselection.
% and for a fixed circuit we estimate the number of samples (i.e. trajectories) scales as $\mathrm{poly}(1/\varepsilon)$, where $\varepsilon$ is the desired precision in $\chi$.

By choosing the circuit bulk to be composed of Clifford operations and $\sigma$ to be a stabilizer state, the protocol is scalable on both the quantum and the classical sides.
Nevertheless, unless $\rho$ is also a stabilizer state, the $\rho$-circuit output state is still highly nontrivial and hard to represent classically.
More broadly, our protocol represents a general -- {although not always scalable} -- approach for experimental observations of measurement-induced physics that does not reduce the quantum simulation to a mere confirmation of a classical computation, see recent examples in Refs.~\cite{garratt2022measurements, feng2022measurementinduced, weinstein2023nonlocality, barratt2022learnability}.

In~\cite{SM}
% we further consider a layer of complete measurement on the output state,
we consider one nontrivial aspect of the output state in the volume law phase when the $\rho$-circuit is not efficiently classically simulable, namely the bistring distribution when all qubits are measured, and found qualitative differences from the Porter-Thomas distribution.

% and show that with a generic (non-stabilizer) choice of $\rho$ the probability distribution over the output bitstrings obeys a nontrivial distribution with a long tail, similar to, but different in detail than, the Porter-Thomas distribution from purely unitary random circuits.
% Straightforwardly sampling from this distribution on a quantum processor still requires postselection, but ways to avoid this may exist. 
% Assuming classical hardness of sampling from this distribution, we discuss the possibility of demonstrating quantum advantage in the hybrid circuit setting.
% We discuss possible implications of this result.

% \section{Linear cross entropy and order parameter \label{sec:results}}

%------------------------------
%------------------------------

\emph{Linear cross entropy and order parameter.} ---
We consider the ``hybrid'' circuit shown in Fig.~\ref{fig:circuit},
composed of unitary gates on nearest-neighbor qubits arranged in a brickwall structure, and single-site measurements in the bulk, performed with probability $p$ at each qubit within each time step.
By convention, each time step contains $L/2$ unitary gates.
Different from the usual setup~\cite{li2019hybrid}, we have an additional ``encoding'' stage before the hybrid evolution for time $t_{\rm encoding} = 2L$, following Refs.~\cite{gullans2019purification, gullans2019scalable}.
The reason for this somewhat unusual choice is practical, to get a clearer experimental signal of the MIPT~\cite{SM}.
We call the evolution after the encoding stage the ``circuit bulk'', which lasts for another $t_{\rm bulk} = 2L$.
The total circuit time is $T = t_{\rm encoding} + t_{\rm bulk} = 4L$.

For concreteness, we take all the measurements to be in the Pauli $Z$ basis.
Given a circuit layout (as determined by the brickwork structure and the location of measurements) and the unitary gates in the bulk -- which we denote collectively as $C$ -- the \emph{unnormalized} output state is defined by $C$ and the measurement record $\bs{m} = \{m_1, m_2, \ldots, m_N\}$ as
\begin{align}
\label{eq:def_rho_m}
    \rho_{\bs{m}} = C_{\bs{m}} \rho C_{\bs{m}}^\dg,
\end{align}
where $\rho$ is the initial state of the circuit, and $C_{\bs{m}}$ is the time-ordered product of all the unitaries and projectors in the circuit, written schematically as
\begin{align}
\label{eq:C_m_time_ordered_expansion}
C_{\bs{m}}
    = &\ P_{m_N} P_{m_{N-1}}
    \ldots 
    P_{m_{N-N_T+1}} \cdot U_T \nn
    & \cdot
    P_{m_{N-N_T}}
    % P_{m_{N-N_T-1}}
    \ldots 
    P_{m_{N-N_T-N_{T-1}+1}} \cdot U_{T-1} \nn
    & \cdot
    P_{m_{N-N_T-N_{T-1}}}
    % P_{m_{N-N_T-N_{T-1}-1}}
    \ldots 
    P_{m_{N-N_T-N_{T-1}-N_{T-2}+1}} \cdot U_{T-2} \nn
    & \ldots
\end{align}
Here each line contains all quantum operations in one circuit time step, and $N$ is the total number of measurements, which is proportional to the spacetime volume of the circuit, $N \propto p V = p L T$.
The corresponding probability of obtaining $\bs{m}$
is given by
% \MF{****Might we want $p_{\bs{m}}^\rho$ to be replaced by $p_{\bs{m}}^\rho$, and similarly for $p_{\bs{m}}^\sigma$, $p_{\bs{m}}^\sigma$, since then only need keep track of $\rho,\sigma$****}
\begin{align}
    p_{\bs{m}}^\rho = \tr \rho_{\bs{m}}.
\end{align}
We define similar quantities for a different initial state $\sigma$,
\begin{align}
    \sigma_{\bs{m}} =&\  C_{\bs{m}} \sigma C_{\bs{m}}^\dg,\\
    p_{\bs{m}}^\sigma =&\ \tr \sigma_{\bs{m}}.
\end{align}
With these, we define the (normalized) linear cross entropy of the circuit between the two initial states as
\begin{align}
\label{eq:chi_C}
    \chi_C = \frac{\sum_{\bs{m}} p_{\bs{m}}^\rho p_{\bs{m}}^\sigma}
    {\sum_{\bs{m}} \(p_{\bs{m}}^\sigma\)^2}.
\end{align}
Here, for fixed choices of $\rho$ and $\sigma$, after averaging over $\bs{m}$, $\chi_C$ only depends on the circuit $C$, and we have explicitly included this dependence in our notation (while keeping the dependence on $\rho$ and $\sigma$ implicit).
Finally, we take its average over $C$,
\begin{align}
\label{eq:chi_average_over_U}
    \chi \coloneqq \mathbb{E}_C \chi_C = \mathbb{E}_C \frac{\sum_{\bs{m}} p_{\bs{m}}^\rho p_{\bs{m}}^\sigma}
    {\sum_{\bs{m}} \(p_{\bs{m}}^\sigma\)^2}.
\end{align}
It was previously pointed out~\cite{choi2019spin} that a quantity closely related to $- \ln \chi$ corresponds to the free energy cost after fixing a boundary condition in a (replicated) spin model~\cite{nahum2018operator, zhou2018emergent, andreas2019hybrid, choi2019spin}; in~\cite{SM},
we provide a similar calculation for our circuit. % with the ``encoding'' stage. \YZ{This sounds like the encoding stage is not relevent for the main text. For the main result the encoding stage is always present, and we should include them in the definition of $C_{\bs{m}}$.}
From this derivation we expect $1-\chi= e^{-O(L)}$ for large $L$ in the volume law phase $(p<p_c)$, and $1-\chi > 0$ in the area law phase $(p>p_c)$, even as $L \rightarrow \infty$.

The physical meaning of $\chi$ %should be 
is clear: it quantifies the difference between the probability distributions over measurement histories for the two initial states.
In the volume law phase, $\chi = 1$ implies the impossibility of distinguishing different initial states from bulk measurements, due to the ``coding'' properties of this phase (i.e. the dynamics in the volume law phase generates a ``dynamical quantum memory''~\cite{gullans2019purification, choi2019qec, fan2020selforganized, li2020capillary, fidkowski2020forget, yoshida2021decoding}).
Intuitively, in the volume law phase, local measurements are so infrequent that it extracts little information about the inital state, as the information is sufficiently scrambled by the random unitaries.
The code breaks down when $p$ is increased past the transition, and $\chi$ saturates to a finite, nonuniversal constant strictly smaller than $1$.
In this phase, information about the initial state leaks into the measurement outcomes.

%--------------------------------
\begin{figure*}[t]
    % \centering
    % \includegraphics[width=1.0\textwidth]{clifford_numerics_main_v2}
    \includegraphics[width=1.0\textwidth]{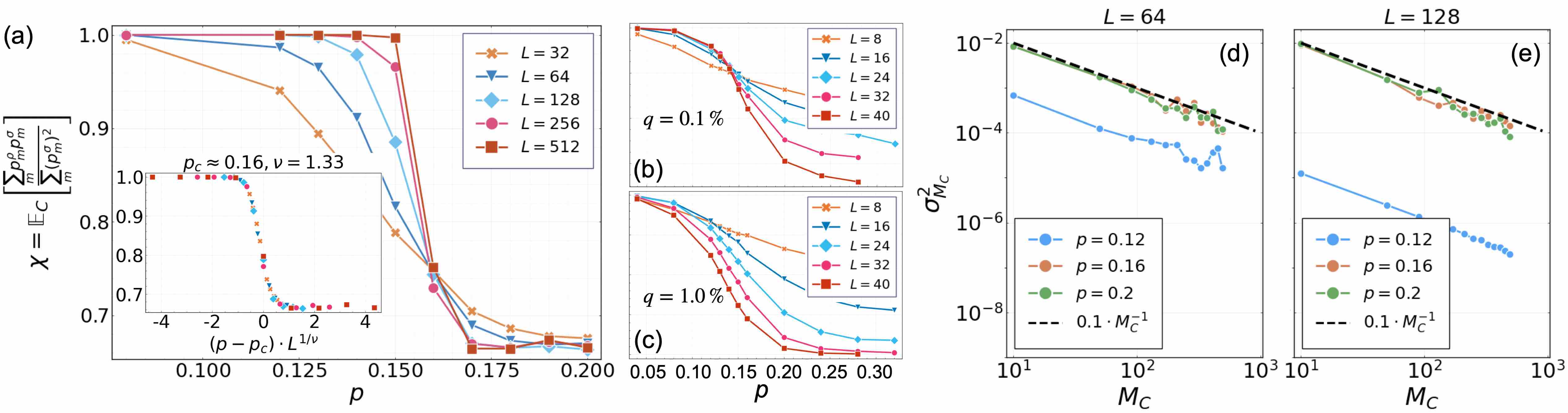}
    \caption{
    (a)
    Numerical results for 
    $\chi_C$
    when averaged over $300$ Clifford circuits in the bulk (denoted by $\mathbb{E}_C$), with the initial states $\rho = \frac{1}{2^L}\mathbb{1}$ and $\sigma = (\ket{0}\bra{0})^{\otimes L}$.
    Here, for each $C$, the calculation is exact, and $M$ can be thought of as infinity in Eq.~\eqref{eq:chi_C_clifford_sample}.
    (Inset) Collapsing the data to a scaling form, with parameters $p_c$ and $\nu$ close to those found near the MIPT in entanglement entropy~\cite{li2018hybrid, li2019hybrid}.
    (b,c) The bahavior of $\chi$ when depolarizing noise is present in the $\rho$-circuit.
    As we see, at noise rate $q = 0.1\%$ (b), there is still evidence for a phase transition, although the location of the transition has shifted from $p_c \approx 0.16$ to $p_c \approx 0.14$.
    At noise rate $1\%$ (c),
    % the phase transition becomes a smooth crossover 
    there is no crossing, and any signature of the phase transition is completely washed out.
    (d, e) The convergence of the sample average $\mu_{M_C} = (M_C)^{-1} \sum_{j=1}^{M_C} \chi_{C_j}$ to $\chi = \mathbb{E}_C \chi_C$ with increasing number of circuit samples $M_C$.
    For each of $L \in \{64, 128\}$, we plot $ \sigma^2_{M_C} = \mathbb{E} \left[ \left(\mu_{M_C} - \chi\right)^2 \right]$ for $M_C \leq 500$, whereas $\chi$ is estimated using $M_C = 2000$ circuits.
    The results are consistent with the central limit theorem, see Eq.~\eqref{eq:sample_variance_chi_C_CLT}.
    From this plot we see that sample variance is suppressed by large $L$ when $p < p_c$, and is independent of $L$ when $p \ge p_c$.
    This justifies our choice of a relatively small $M_C$ that is independent of the system size.
    }
    \label{fig:chi_clifford_rho_stab_sigma_stab}
\end{figure*}
%--------------------------------

% \subsection{Generic hardness of estimating $\chi$ \label{sec:generic_hardness}}

% \EA{I'm not sure this should be a separate section. I'm also not sure the hardness explained here is the most crucial point. I think the main thing to say is that if neither $\sigma$ or $\rho$ can be simulated efficiently, then taking the overlap between the two distributions is hard (requires reconstructing the distributions). On the other hand if $\sigma$ is a stabilizer state we can calculate the $O(1)$ numbers $p^\sigma_m/\sum_m (p^\sigma_m)^2$ and sample over them }

We now outline a protocol for estimating $\chi$, which is similar to the linear cross entropy benchmark (``linear XEB'') for random unitary circuits~\cite{boixo2016XEB,google2019supremacy}.
Then we discuss its limitations when applied to the MIPT and how to overcome them in case of a stabilizer circuit.

% \subsection{General setup}
\emph{General setup}.---
Consider running the circuit with initial state $\rho$ (``the $\rho$-circuit'') on a quantum simulator.
From the simulation we obtain a measurement record $\bs{m}$, an event that occurs with probability $p_{\bs{m}}^\rho$.
Given $\bs{m}$ we can perform a classical simulation with the initial state $\sigma$, and calculate the corresponding probablity $p_{\bs{m}}^\sigma$.
Repeating this $M$ times, we obtain a sequence of probabilities $\{p_{\bs{m}_1}^\sigma, p_{\bs{m}_2}^\sigma, \ldots p_{\bs{m}_M}^\sigma \}$.
% \st{We expect their mean to converge}
Their mean converges to the numerator of Eq.~\eqref{eq:chi_C}, 
\begin{align}
\label{eq:chi_C_numerator_estimator}
\lim_{M \to \infty} \avg{p^{\sigma}_{\bs{m}_{j=1}^M}}_\rho \coloneqq
\lim_{M \to \infty} \frac{1}{M} \sum_{j=1}^{M} p^{\sigma}_{\bs{m}_{j}} = 
\sum_{\bs{m}} p_{\bs{m}}^\rho p_{\bs{m}}^\sigma .
\end{align}

The denominator of Eq.~\eqref{eq:chi_C} can be estimated similarly with a separate classical simulation, by running the $\sigma$-circuit $M'$ times, and computing the mean of probabilities $\{p^{\sigma}_{\bs{m}_j}\}$.
This way we get
\begin{align}
\label{eq:chi_C_denominator_estimator}
\lim_{M' \to \infty} \avg{p^{\sigma}_{\bs{m}_{j=1}^{M'}}}_\sigma
\coloneqq
\lim_{M' \to \infty} \frac{1}{M'} \sum_{j=1}^{M'} p^{\sigma}_{\bs{m}_j} = 
\sum_{\bs{m}} \(p_{\bs{m}}^\sigma\)^2.
\end{align}
Both equations above are well-defined, and in this protocol each run of the circuit is used, so no postselection is required.
This should lead to a general protocol for experimentally probe MIPTs, although a full classical simulation is still necessary, and the experimentally accessible system size will be limited by the power of classical simulation. %see Sec.~\ref{sec:random_haar_numerics}.

To obtain a scalable protocol, we first focus on the case where $\sigma$ is a stabilizer state, and the circuit bulk $C_{\bs{m}}$ is composed of stabilizer operations (Clifford gates and Pauli measurements)~\cite{gottesman1997thesis, gottesman1998heisenberg, aaronson2004chp}. 
At this point we do not put constraint on $\rho$.
% As we explain, this can greatly
In this special case, the denominator of Eq.~\eqref{eq:chi_C} can be computed exactly in polynomial time, without doing any sampling as in Eq.~\eqref{eq:chi_C_denominator_estimator}~\cite{SM}.
Thus, we may rewrite  Eq.~\eqref{eq:chi_C} as
\begin{align}
    \label{eq:chi_C_clifford_rewrite}
    \chi_C =  \sum_{\bs{m}} p_{\bs{m}}^\rho \frac{p_{\bs{m}}^\sigma}
    {\sum_{\bs{m}} \(p_{\bs{m}}^\sigma\)^2},
\end{align}
and in analogy with Eq.~\eqref{eq:chi_C_numerator_estimator}, 
\begin{align}
    \label{eq:chi_C_clifford_sample}
    \chi_C = \lim_{M \to \infty} \avg{ \frac{p^\sigma_{\bs{m}_{j=1}^M}} {\sum_{\bs{m}} \(p_{\bs{m}}^\sigma\)^2}
    }_\rho.
\end{align}
For each run of the $\rho$-circuit, we obtain the measurement record $\bs{m}_j$ and compute $\frac{p^\sigma_{\bs{m}_j}} {\sum_{\bs{m}} \(p_{\bs{m}}^\sigma\)^2}$ in polynomial time, and take its mean over runs.
Since the circuit is Clifford,
the new ``observable'' $\frac{p^\sigma_{\bs{m}_j}} {\sum_{\bs{m}} \(p_{\bs{m}}^\sigma\)^2}$ is either $0$ or $1$ for a given $\bs{m}$,\footnote{\label{fn:Nrand}Recall that for Cliffford circuits a measurement either has a deterministic outcome, or has random outcomes $\pm 1$ with equal probabilities $1/2$~\cite{aaronson2004chp}.
Let $N_{\rm rand}$ be the number of measurements (out of the total $N$) whose outcome is randomly $\pm 1$.
There are $2^{N_{\rm rand}}$ possible trajectories in total, and they occur with equal probabilities $p^\sigma_{\mathbf{m}} = 2^{-N_{\rm rand}}$.
See Supplemental Material~\cite{SM} for more details.
}
and this average converges quickly with increasing $M$.
In particular, since this is a binary random variable, the variance of the samples should decay as $M^{-1/2}$ for a given $C$.
Thus, for a fixed circuit $M$ scales as
% % $\mathrm{poly}(1/\varepsilon)$.
$1/\varepsilon^2$, where $\varepsilon$ is the error of the estimation of $\chi_C$.
We also see that $\chi_C$ is always bounded between $0$ and $1$.
This is a property special to Clifford circuits.

\emph{Numerical methods and results}. ---
We first take $\rho$ to be a stabilizer state, while keeping $\sigma$ another stabilizer state.
% When this condition is satisfied a further simplification occurs, 
As we explain in~\cite{SM}
% . In particular, 
now $\chi_C$ in Eq.~\eqref{eq:chi_C_clifford_rewrite} admits a closed form expression that does not involve any summation over $\bs{m}$.
% see Eq.~\eqref{eq:chi_C_stabilizer_register}. 
This allows an exact calculation of $\chi_C$ without the need of performing any sampling, at the cost of introducing $N$ extra qubits that record the measurement history.
These qubits are usually called ``registers''.

A further simplification occurs when $\rho$ is obtainable from $\sigma$ via erasure or dephasing channels, so that the $N$ register qubits can also be dispensed with~\cite{SM}.
% ; see Eq.~\eqref{eq:chi_C_Nrand}.
We will focus on this case below, where the numerical simulation is most scalable so that we can confidently extrapolate the results to more general choices of $\rho$.

In Fig.~\ref{fig:chi_clifford_rho_stab_sigma_stab}(a), we plot $\chi = \mathbb{E}_C \chi_C$ for $\rho = \frac{1}{2^L} \mathbb{1}$ and $\sigma = (\ket{0}\bra{0})^{\otimes L}$, which satisfies the condition above.
% \st{computed using the exact numerical method.}
The data shows a clear ``crossing'' of $\chi$ near the transition, confirming our expectation that $\chi$ is an order parameter for the MIPT.
Indeed, in the large $L$ limit and for $p<p_c$, $\chi$ approaches unity, demonstrating that the distributions of measurement outcomes become equal, independent of the initial state.
Moreover,
data collapse in Fig.~\ref{fig:chi_clifford_rho_stab_sigma_stab}(d) shows good agreement to a standard scaling form, with numerical values of the location of the transition $p_c$ and of the critical exponent $\nu$ close to previous characterizations of the MIPT~\cite{li2019hybrid}.

\begin{figure}[t]
    % \centering
    % \includegraphics[width=.45\textwidth]{./figs/chi_clifford_rho_magic_sigma_stab}
    % \includegraphics[width=.45\textwidth]{figs/chi_haar.jpeg}
    \includegraphics[width=.45\textwidth]{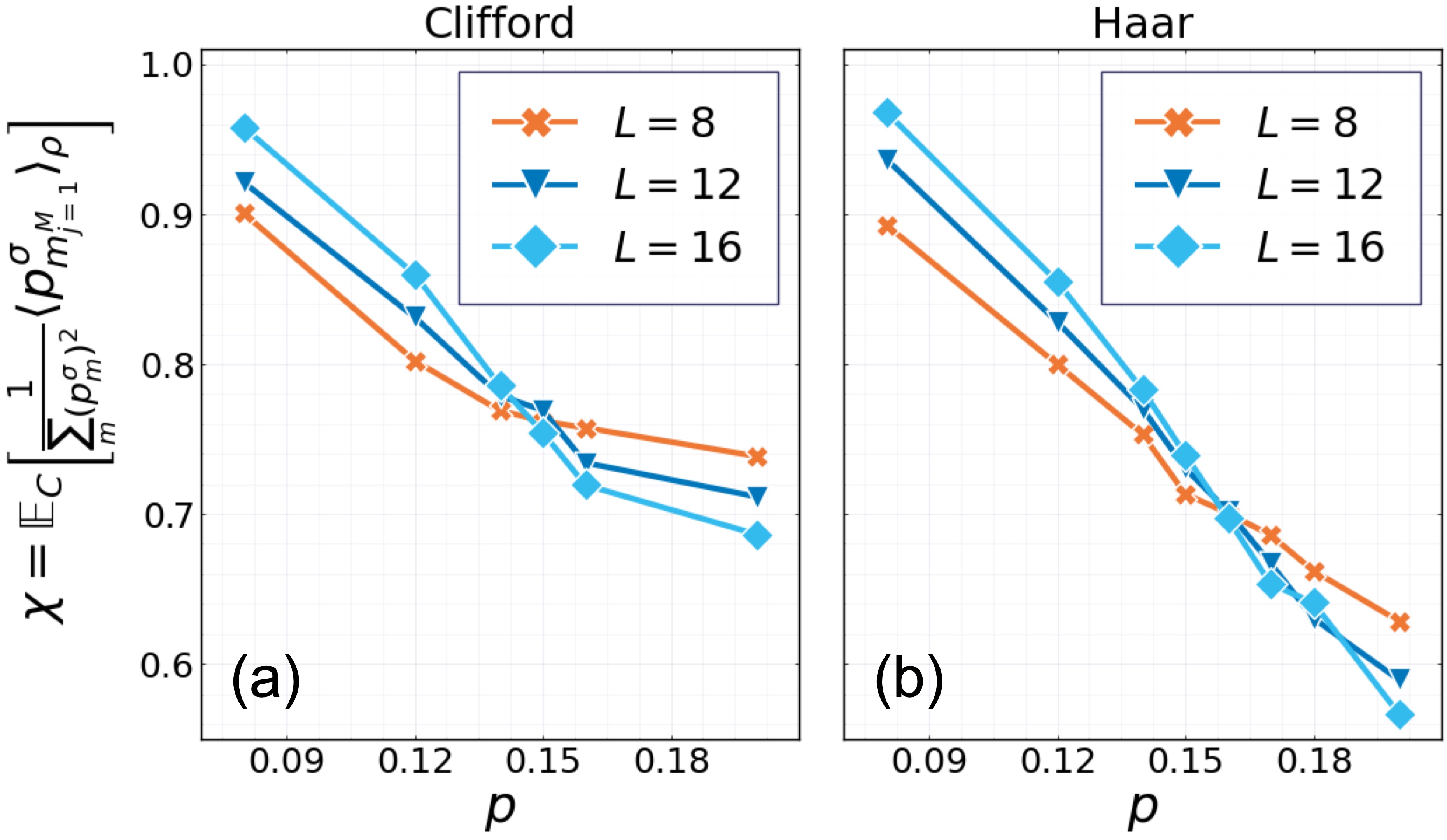}
    \caption{
    (a) 
    Numerical results of $\chi$ for initial states $\rho = \bigotimes_{i=1}^{L/2} (\ket{0}\bra{0}_{2i-1} \otimes \ket{T}\bra{T}_{2i})$ and $\sigma =(\ket{0}\bra{0})^{\otimes L}$
    obtained from random Clifford circuits.
    Here, $M_C = 300$ circuit realizations are taken for each $L$, %, and for each circuit $M = 100$ shots of the circuit are taken.
    and for each circuit, we use $M = 100$ runs to estimate $\chi_C$, following  Eq.~\eqref{eq:chi_C_clifford_sample}.
    Compared to Fig.~\ref{fig:chi_clifford_rho_stab_sigma_stab}(a), the results are qualitatively similar,
    despite a different choice of initial state and smaller system sizes.
    (b) 
    Numerical results of $\chi$ for initial states $\rho = (\ket{+}\bra{+})^{\otimes L}$ and $\sigma =(\ket{0}\bra{0})^{\otimes L}$ obtained from random Haar circuits.
    Here, $M_C = 150$ circuit realizations are taken for each $L$, and for each circuit we estimate Eq.~\eqref{eq:chi_C_numerator_estimator} and Eq.~\eqref{eq:chi_C_denominator_estimator} separately, using $M = 3200$ runs each.
    }
    \label{fig:chi_clifford_rho_magic_sigma_stab}
\end{figure}

An important practical parameter is the number $M_C$\footnote{Notice that this is \textit{the number of samples of $C$}, and is different from $M$ as discussed above, which \textit{the number of runs (or trajectories) taken for each circuit sample $C$}.
In this discussion we assume that $\left\{ \chi_{C_{j=1}^{M_C}} \right\}$ have been obtained following previous discussions around Eq.~\eqref{eq:chi_C_clifford_sample}, when $M$ is taken to infinity.} of circuit samples needed to estimate $\chi$ within a given accuracy, in particular their scaling with the system size.
By the central limit theorem, given independent samples $\{C_{j=1}^{M_C}\}$,
the sample average $\mu_{M_C} = \frac{1}{M_C} \sum_{j=1}^{M_C} \chi_{C_j}$ converges to $\chi$ at large $M_C$ as follows, 
\begin{align}
\label{eq:sample_variance_chi_C_CLT}
    \sigma^2_{M_C} = \mathbb{E} \left[ \left(\mu_{M_C} - \chi\right)^2 \right] \propto (M_C)^{-1},
\end{align}
with an overall amplitude that converges to the variance of $\chi_C$, $\sigma^2[\chi_C] \coloneqq \mathbb{E}_C[\chi_C^2] - (\mathbb{E}_C[\chi_C])^2$.
In Fig.~\ref{fig:chi_clifford_rho_stab_sigma_stab}(d,e) we compute $\sigma^2_{M_C}$ numerically at two different system sizes $L$ and at different locations of the phase diagram.
Our results confirm Eq.~\eqref{eq:sample_variance_chi_C_CLT}, and by fitting the overall amplitude we find that $\sigma^2[\chi_C]$ is suppressed by large $L$ in the volume law phase $p<p_c$ (as consistent with $\chi \to 1$), and saturates to an $L$-independent constant ($\approx 0.1$) for $p \geq p_c$.
Together with our previous discussion on $M$ (number of runs per circuit $C$), these results justify our choices of relatively small $M_C$ and $M$ that are independent of system sizes, see Fig.~\ref{fig:chi_clifford_rho_stab_sigma_stab} and Fig.~\ref{fig:chi_clifford_rho_magic_sigma_stab} (a) below.

We also consider the effect of depolarizing noise, occuring randomly in the $\rho$-circuit with probability $q$ per qubit per time step; whereas the $\sigma$-circuit is still taken to be noiseless.
The setup is to mimic an experimental sampling procedure, where we run the $\rho$-circuit on a quantum processor subject to noise, whereas our supplemental classical simulation of the $\sigma$-circuit is noiseless.
The depolarizing noise acts as a symmetry-breaking field in the effective spin model~\cite{andreas2019hybrid, choi2019spin, li2020cft, li2021dpre, zhangpengfei2021magneticfield, ippoliti2020postselectionfree, ippoliti2021fractal},\footnote{See also Refs.~\cite{noh2020efficient, deshpande2021tight, dalzell2021random} for related discusssion in random unitary circuits.} and in its presence the MIPT is no longer sharply defined.
Nevertheless, evidence of the MIPT may still be observable if the error rate is small compared to the inverse spacetime volume of the circuit, as we see in 
% Fig.~\ref{fig:chi_clifford_rho_stab_sigma_stab_depolarizing}.
Fig.~\ref{fig:chi_clifford_rho_stab_sigma_stab}(b,c). 
% \YL{We may further analyze this at the critical point and extract the exponents $\delta$ and $\gamma$.}

% \subsubsection{Magic state $\rho$ versus stabilizer state $\sigma$}

Next, we take $\rho$ to be a non-stabilizer state, and $\sigma$ to be a stabilizer state.
In particular, we choose a state with $\ket{0}$ and $\ket{T}$ on alternating sites,
\begin{align}
\label{eq:rho_T_gate_half_L}
    \rho = \bigotimes_{i=1}^{L/2} (\ket{0}\bra{0}_{2i-1} \otimes \ket{T}\bra{T}_{2i}),
% ^{\otimes L/2},
\end{align}
where $\ket{T} = \frac{1}{\sqrt{2}}\( \ket{0}+e^{i\pi/4}\ket{1}\)$ is a magic state.
We still take the other initial state to be $\sigma = (\ket{0}\bra{0})^{\otimes L}$.
% \YL{}

%--------------------------
%--------------------------

Based on our calculations~\cite{SM}, we expect $\chi_C$ to exhibit similar behavior as in Fig.~\ref{fig:chi_clifford_rho_stab_sigma_stab}.
This is confirmed in Fig.~\ref{fig:chi_clifford_rho_magic_sigma_stab}(a), where we follow the sampling procedure in Eq.~\eqref{eq:chi_C_clifford_sample}.
In particular, for a given $C$, we take $L \in \{8, 12, 16\}$, and sample $M = 100$ measurement trajectories, and compute
$
    \avg{ \frac{p^\sigma_{\bs{m}_{j=1}^M}} {\sum_{\bs{m}} \(p_{\bs{m}}^\sigma\)^2}
    }_\rho \approx \chi_C.
$
We then take the average over many different choices of $C$, namely $\mathbb{E}_C \avg{ \frac{p^\sigma_{\bs{m}_{j=1}^M}} {\sum_{\bs{m}} \(p_{\bs{m}}^\sigma\)^2}}_\rho \approx \mathbb{E}_C \chi_C$.
We observe a crossing of $\chi$ at roughly the same value of $p_c$ in Fig.~\ref{fig:chi_clifford_rho_stab_sigma_stab}(a).
% \st{ These results are comparable to those in Fig.~\ref{fig:chi_clifford_rho_stab_sigma_stab}(a). }
% It is important to  notice that though the classical side of the computation (the $\sigma$-circuit) can be carried out efficiently, the quantum side (the $\rho$-circuit) is still classically hard~\cite{bravyi2016improved}.
The system sizes that we accessed are limited by classical simulations of the $\rho$-circuit~\cite{bravyi2016improved}, but we hope larger system sizes can be achieved on near-term quantum processors.

% \subsubsection{Numerical results in random Haar circuits \label{sec:random_haar_numerics}}

% %------------------------------------
% \begin{figure}
%     \centering
%     \includegraphics[width=.23\textwidth]{figs/chi_haar.jpeg}
%     \caption{Caption}
%     \label{fig:chi_haar}
% \end{figure}
% %------------------------------------

Finally, to test the validity of our approach beyond Clifford circuits, we calculate $\chi$ in circuits with random Haar unitary gates, for  $\rho = (\ket{+}\bra{+})^{\otimes L}$ and $\sigma = (\ket{0}\bra{0})^{\otimes L}$.
Here we have to estimate the normalization of $\chi_C$ (see Eqs.~(\ref{eq:chi_C}, \ref{eq:chi_C_denominator_estimator})) separately.
To obtain plots with comparable accuracy as those from Clifford circuits,
the number of runs per circuit needs to be at least an order of magnitude larger (for system sizes up to $L=16$), due to the additional numerical uncertainty in the normalization.
Our results are shown in
Fig.~\ref{fig:chi_clifford_rho_magic_sigma_stab}(b), with an overall trend consistent with a phase transition.
% where we observe a crossing of $\chi$ at roughly the same value of $p_c$

% \section{Discussions}
\emph{Discussions}. ---
Our protocol requires a simulation of many instances of the random hybrid circuit with mid-circuit measurements, and for each instance $O(1/\varepsilon^2)$ trajectories to estimate the cross entropy to accuracy $\varepsilon$.
This should be a task of similar complexity to Google's simulation of random unitary circuits~\cite{google2019supremacy}, except that here we do not make measurements on the output state but in the bulk.
However, different from that experiment, for observing the MIPT it suffices to focus on Clifford circuits, for which the classical simulation is not hard.
This protocol is thus as scalable as the quantum processors.
Our protocol does not require extra quantum operations, and is flexible in the choice of the initial state.
The signal for the phase transition persists at $L = 40$ for sufficiently weak ($\approx 0.1\%$) depolarizing noise.
Thus, we hope this protocol might be achievable on existing or near-term devices.

% \YL{Experimental prospects?}

If the circuit is not composed of Clifford gates, our protocol is expected to require exponential classical resources.
It is presently unclear whether it is in fact possible to probe the MIPT beyond Clifford circuits with polynomial resources~\cite{dehghani2022neuralnetwork}.

Although the classical simulation is chosen to be easy for practical purposes, in our protocol the quantum simulation is classically hard for a generic choice of the initial state, which would result in a highly nontrivial output state.
Our numerical results in~\cite{SM} suggest that sampling measurement outcomes on the output state of the quantum simulation  is classically hard in the volume law phase.
Whether this can be used in practice for demonstrating quantum advantage is not known, due to apparent need of postselection in order to sample from this distribution.

\emph{Acknowledgements}.---
We acknowledge  helpful discussions with Tanvi Gujarati, Jacob Hauser, Hirsh Kamakari, Vedika Khemani, Jin Ming Koh, Ali Lavasani, Austin Minnich, Mario Motta, Alan Morningstar, Xiao-Liang Qi, Shi-Ning Sun, Shengqi Sang, Jonathan Thio, Sagar Vijay, and Sisi Zhou.
We thank Michael Gullans and Edward Chen for useful suggestions, and Matteo Ippoliti for pointing out a mistake in an earlier version of the Supplemental Material~\cite{SM}.
YL is grateful for the hospitality of Vedika Khemani at Stanford University, where much of this work was undertaken.
This work was supported by the Heising-Simons Foundation (YL and MPAF),
and by the Simons Collaboration on Ultra-Quantum Matter, which is a grant from the Simons Foundation (651457, MPAF).
YZ is supported by the Q-FARM fellowship at Stanford University. 
PG is supported by the Alfred P. Sloan Foundation through Grant FG-2020-13615, the Department of Energy through Award DE-SC0019380, and the Simons Foundation through Award No. 620869.
EA is supported in part by the NSF QLCI program through grant number OMA-2016245.
Use was made of computational facilities purchased with funds from the National Science Foundation (CNS-1725797) and administered by the Center for Scientific Computing (CSC). The CSC is supported by the California NanoSystems Institute and the Materials Research Science and Engineering Center (MRSEC; NSF DMR-1720256) at UC Santa Barbara.

\let\oldaddcontentsline\addcontentsline% Store \addcontentsline
\renewcommand{\addcontentsline}[3]{}% Make \addcontentsline a no-op
\bibliography{refs}
\let\addcontentsline\oldaddcontentsline% Restore \addcontentsline

\clearpage
% \appendix

\setcounter{equation}{0}
\setcounter{figure}{0}
\setcounter{table}{0}
\setcounter{page}{1}
\makeatletter
\renewcommand{\thesection}{S\arabic{section}}
\renewcommand{\theequation}{S\arabic{equation}}
\renewcommand{\thefigure}{S\arabic{figure}}
% \renewcommand{\bibnumfmt}[1]{[S#1]}

%---------------------------------------------------------

% \preprint{arXiv:xxxx.xxxxx[cont-mat]}
% \appendix 

\begin{widetext}
\begin{center}
\textbf{\large Supplemental Material for ``Cross Entropy Benchmark for Measurement-Induced Phase Transitions''}
\\~\\
Yaodong Li$^{1,2}$, Yijian Zou$^{2}$, Paolo Glorioso$^{2}$, Ehud Altman$^{3}$, Matthew P. A. Fisher$^{1}$ \\
\vspace{.05in}
\small{
$^{1}$\textit{Department of Physics, University of California, Santa Barbara, CA 93106}\\
$^{2}$\textit{Department of Physics, Stanford University, Stanford, CA 94305}\\
$^{3}$\textit{Department of Physics, University of California, Berkeley, CA 94720}
}
\end{center}
\end{widetext}

\tableofcontents

\section{Cross entropy as boundary correlation function \label{sec:map_chi_spin}}

\subsection{Bulk cross entropy with encoding \label{sec:chi_bulk}}

%----------------------------
\begin{figure*}[t]
    \centering
    \includegraphics[width=\textwidth]{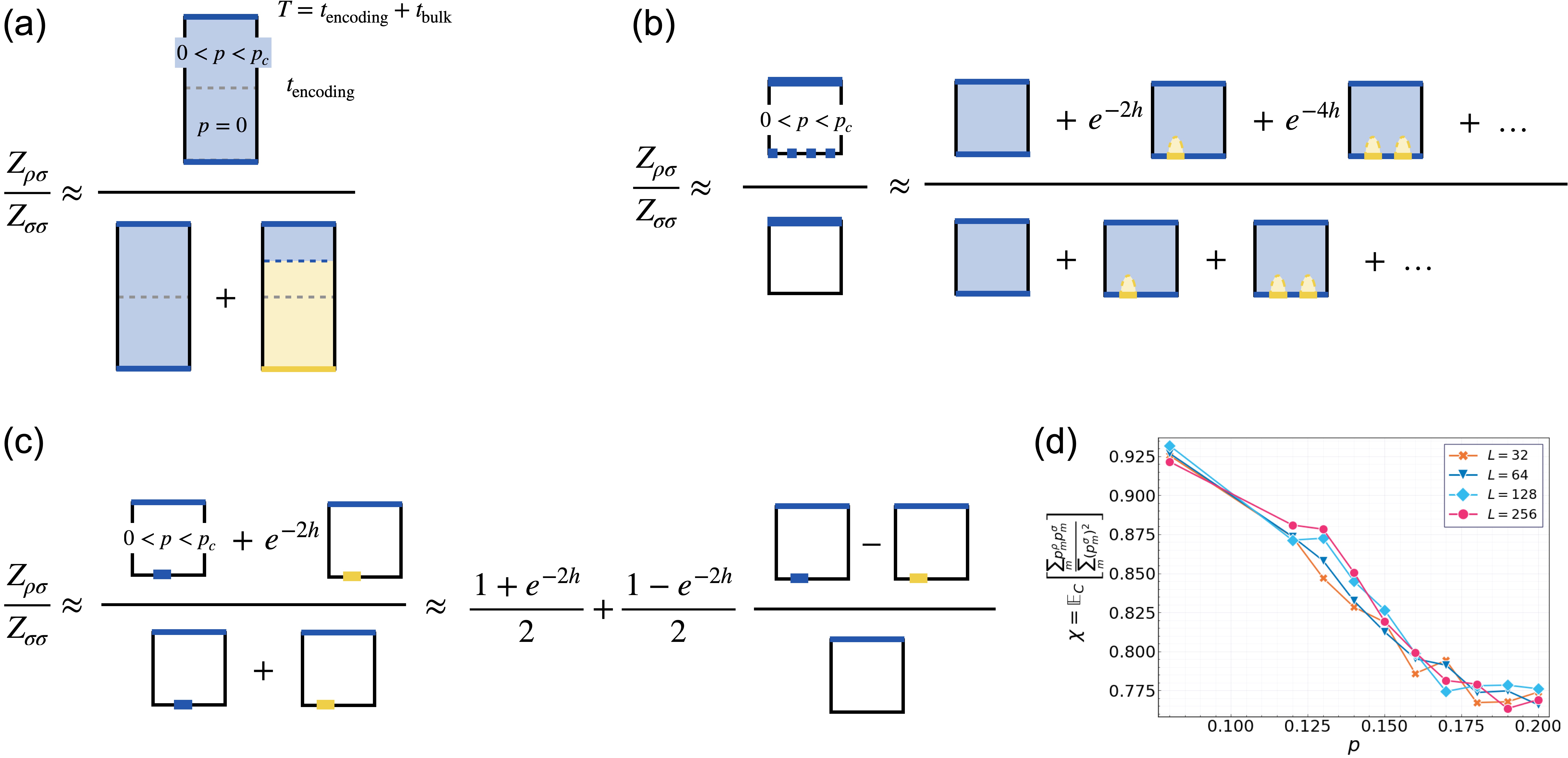}
    % \quad\quad\quad
    % \includegraphics[height=.23\textwidth]{figs/ZoverZ_noencode}
    % \includegraphics[height=.23\textwidth]{figs/ZoverZ_FXEB}
    \caption{
    (a) 
    Pictorial representation of the partition function ratio $Z_{\rho\sigma} / Z_{\sigma\sigma}$ in Eq.~\eqref{eq:chi_ZoverZ}, for $p=0$ in the encoding stage and $p < p_c$ in the circuit bulk.
    (b)
    Pictorial representation of the partition function ratio in Eq.~\eqref{eq:ZoverZ_noencode_uniform_field}.
    Here we do not have an encoding stage, and there is a \textit{uniform}, finite magnetic field  of strength $h$ (represented with a dashed line) applied at the $t=0$ boundary.
    (c)
    Pictorial representation of the partition function ratio in Eq.~\eqref{eq:ZoverZ_noencode_local_field}.
    Here we do not have an encoding stage, and there is a \textit{local}, finite magnetic field of strength $h$ applied at the $t=0$ boundary.
    In this case, the cross entropy is expected to be a function $p$ but not of $L$ (see Eq.~\eqref{eq:ZoverZ_noencode_local_field_result}), as we confirm in (d).
    In all figures 
    the blue color represents spins pointing in the ``$+$'' direction, the yellow color represents spins pointing in the ``$-$'' direction, and the black color represents a ``free'' boundary condition, where the spins can point in either direction.
    }
    \label{fig:ZoverZ}
\end{figure*}
%----------------------------

We unpack the circuit averaged linear cross entropy $\chi$ defined in Eq.~\eqref{eq:chi_average_over_U},
\begin{align}
    \chi \coloneqq &\ \mathbb{E}_C \chi_C \nn
    =&\ \mathbb{E}_C \frac{\sum_{\bs{m}} p_{\bs{m}}^\rho p_{\bs{m}}^\sigma}
    {\sum_{\bs{m}} \(p_{\bs{m}}^\sigma\)^2} \nn
    =&\ \mathbb{E}_C \frac{\sum_{\bs{m}} \(\tr C_{\bs{m}} \rho C_{\bs{m}}^\dg \)
    \(\tr C_{\bs{m}} \sigma C_{\bs{m}}^\dg \)}
    {\sum_{\bs{m}} 
    \(\tr C_{\bs{m}} \sigma C_{\bs{m}}^\dg \)^2
    } \nn
    =&\ \mathbb{E}_C
    \frac{
        \sum_{\bs{m}} \tr C_{\bs{m}}^{\otimes 2} \cdot (\rho \otimes \sigma) \cdot C_{\bs{m}}^{\dg\otimes 2}
    }
    {
        \sum_{\bs{m}} \tr C_{\bs{m}}^{\otimes 2} \cdot (\sigma \otimes \sigma) \cdot C_{\bs{m}}^{\dg\otimes 2}
    }.
\end{align}
Recall that the letter $C$ encodes the circuit layout (i.e. the locations of unitary gates and measurements) and the unitary gates, but not the measurement outcomes.
The summation over $\bs{m}$ is taken inside the average, in both the numerator and denominator, independently.
Thus, $\chi$ is different from the trajectory-averaged entanglement entropies that are used previously for identifying the MIPT.
Nevertheless, in Fig.~\ref{fig:chi_clifford_rho_stab_sigma_stab} we see that the location of the transition and the critical exponent $\nu$ do not change much when we use $\chi$ as an order parameter.

A proper treatment of the quenched average leads to a replicated spin model.\footnote{Due to the difference we stressed above, this leads to a stat mech model that differs from those obtained in Refs.~\cite{andreas2019hybrid, choi2019spin}.
In particular, the spins here take values in the permutation group $S_{Q=2n}$ with the replica limit $n \to 0$, and which has a different symmetry.}
For our purposes here, we can instead consider the annealed average~\cite{fan2020selforganized, li2020capillary}, while keeping in mind that this is only a illustrative tool.
In particular, consider 
\begin{align}
\label{eq:chi_annealed}
\overline{\chi} =&\ 
    \frac{
    \mathbb{E}_C
        \sum_{\bs{m}} \tr C_{\bs{m}}^{\otimes 2} \cdot (\rho \otimes \sigma) \cdot C_{\bs{m}}^{\dg\otimes 2}
    }
    {
    \mathbb{E}_C
        \sum_{\bs{m}} \tr C_{\bs{m}}^{\otimes 2} \cdot (\sigma \otimes \sigma) \cdot C_{\bs{m}}^{\dg\otimes 2}
    }.
\end{align}
After the average over $C$, the numerator and the denominator each becomes an Ising partition function on a triangular lattice.
They have bulk weights $J_p(s_i, s_j; s_k)$ for each downward-pointing triangle~\cite{andreas2019hybrid, choi2019spin} (see also Refs.~\cite{nahum2018operator, zhou2018emergent, zhou2019membrane}), and only differ in their boundary conditions.
We denote them $Z_{\rho \sigma}$ and $Z_{\sigma \sigma}$, respectively.

We take $\rho$ and $\sigma$ to be products of local density matrices, i.e.
\begin{align}
\label{eq:rho_sigma_product_states}
    \rho = \prod_{x=1}^L \rho_x, \quad
    \sigma = \prod_{x=1}^L \sigma_x, \text{ where } \tr \rho_x = \tr \sigma_x = 1 \ \forall x.
\end{align}
Moreover, we also have $\tr \sigma_x^2 = 1$ since we assumed $\sigma$ is a pure product state.
Thus,
\begin{widetext}
\begin{align}\label{zss1}
    Z_{\sigma \sigma} =&\ \sum_{\{s_i = \pm 1\}} \prod_{\langle i, j, k \rangle \in \triangledown} J_p(s_i, s_j; s_k) 
    \cdot \prod_{x \in \partial \mc{M}_T} \delta_{s_x=+1}
    \cdot \prod_{x \in \partial \mc{M}_0} \(\delta_{s_x=+1} (\tr \sigma_x)^2 + \delta_{s_x=-1} \tr (\sigma_x^2) \) \nn
    =&\ \sum_{\{s_i = \pm 1\}} \prod_{\langle i, j, k \rangle \in \triangledown} J_p(s_i, s_j; s_k) \cdot \prod_{x \in \partial \mc{M}_T} \delta_{s_x=+1},
\end{align}
and
\begin{align}\label{zrs1}
    Z_{\rho \sigma} =&\ \sum_{\{s_i = \pm 1\}} \prod_{\langle i, j, k \rangle \in \triangledown} J_p(s_i, s_j; s_k) 
    \cdot \prod_{x \in \partial \mc{M}_T} \delta_{s_x=+1}
    \cdot \prod_{x \in \partial \mc{M}_0} \(\delta_{s_x=+1} (\tr \rho_x)(\tr \sigma_x) + \delta_{s_x=-1} \tr (\rho_x \cdot \sigma_x) \) \nn
    =&\ \sum_{\{s_i = \pm 1\}} \prod_{\langle i, j, k \rangle \in \triangledown} J_p(s_i, s_j; s_k) 
    \cdot \prod_{x \in \partial \mc{M}_T} \delta_{s_x=+1}
    \cdot \prod_{x \in \partial \mc{M}_0} \(\delta_{s_x=+1} + \delta_{s_x=-1} \tr (\rho_x \cdot \sigma_x) \) \nn
    =&\ \sum_{\{s_i = \pm 1\}} \prod_{\langle i, j, k \rangle \in \triangledown} J_p(s_i, s_j; s_k)
    \cdot \prod_{x \in \partial \mc{M}_T} \delta_{s_x=+1}
    \cdot
    \prod_{x \in \partial \mc{M}_0}
    e^{h_x (s_x-1)}.
\end{align}
\end{widetext}
Here, we use $\pd \mc{M}_0$ to denote the $t=0$ boundary of the circuit, and $\pd \mc{M}_T$ to denote the final time ($t=T$) boundary.
We see that at $t=0$, $Z_{\sigma\sigma}$ has a ``free'' boundary condition, and $Z_{\rho\sigma}$ has a magnetic field with strength $h_x = -\frac{1}{2} \ln \lz \tr (\rho_x \cdot \sigma_x) \rz$.
At $t=T$, in both partition functions spins are fixed to be $s_x = +1$.

Our circuit in Fig.~\ref{fig:circuit} has an ``encoding'' stage without measurements ($p=0$) up until $t_{\rm encoding} = 2L$.
This makes the lower half of the circuit 
a pure unitary one,
where domain walls with both endpoints on the $t=0$ boundary are disallowed by the microscopics of the stat mech model~\cite{nahum2018operator,zhou2018emergent}.
% deep within the low-temperature ordered phase of the Ising model, and we can essentially assume that the spins are fully aligned, and neglect all thermal fluctuations in the lower half when $L$ is large. \YL{rephrase zero temperature in terms of constraints from the triangular weights.}
In this case, the finite-strength field at the $t=0$ boundary of $Z_{\rho\sigma}$ becomes essentially infinite, putting a hard boundary condition at $t=0$:
\begin{align}
    \label{eq:Z_rho_sigma}
    Z_{\rho \sigma} 
    \approx&\ \sum_{\{s_i = \pm 1\}} \prod_{\langle i, j, k \rangle \in \triangledown} J_p(s_i, s_j; s_k) 
    \nn &\quad \quad
    \cdot \prod_{x \in \partial \mc{M}_T} \delta_{s_x=+1}
    \cdot \prod_{x \in \partial \mc{M}_0} \delta_{s_x=+1}\nn
    \coloneqq&\  Z_{++},
\end{align}
where $Z_{++}$ denotes the partition function with $+$ boundary condition at $t=0$ and $+$ boundary condition at $t=T$.
By the same reasonining and the same notation, we can rewrite
\begin{align}
    \label{eq:Z_sigma_sigma}
    Z_{\sigma \sigma} \approx Z_{++} +  Z_{-+}.
\end{align}

We represent 
these partition functions
diagrammatically in Fig.~\ref{fig:ZoverZ}, where the boundary conditions are highlighted with color: blue for $+$ and orange for $-$.
(In the figure, we only illustrated the case where $p<p_c$ after the initial encoding stage; these two stages are separated by a gray, dashed line.)
We have
\begin{align}
\label{eq:chi_ZoverZ}
    \overline{\chi} = \frac{Z_{\rho\sigma}}{Z_{\sigma\sigma}} = \frac{1}{1 + Z_{-+}/Z_{++}}.
\end{align}
In the volume law phase, we expect  $Z_{-+}/Z_{++} \propto \exp(-O(L))$, because a domain wall with finite line tension of length $L$ must be inserted between $t \in [t_{\rm encoding}, T]$, to accommodate the boundary conditions change from $-$ to $+$ in time; see Fig.~\ref{fig:ZoverZ}.
On the other hand, in the area law phase, the domain wall line tension vanishes, and we have $Z_{-+}/Z_{++} = O(1)$.
Thus, 
\begin{align}
    \overline{\chi} = \begin{cases}
        1 + \exp(-O(L)), & p < p_c \\
        O(1), & p > p_c
    \end{cases}.
\end{align}
Despite the fact that we are adopting an annealed average in $\overline{\chi}$, it captures the qualitative behavior of the quenched average $\chi$ in Eq.~\eqref{eq:chi_average_over_U} in the two phases (but presumably not the critical properties).

\subsection{Bulk cross entropy without encoding \label{sec:chi_bulk_no_encoding}}

Here we briefly discuss the choice of the circuit architecture in Fig.~\ref{fig:circuit}, especially the importance of the encoding stage.
% Here we extend our discussion in Sec.~\ref{sec:why_encoding} on $\chi$ 
Suppose the encoding stage is absent, so that 
the entire two-dimensional magnet is now at finite temperature; see Fig.~\ref{fig:ZoverZ}(b).
Here, the partition functions $Z_{\sigma\sigma}$ and $Z_{\rho\sigma}$ have boundary conditions that are identical to those in Eqs.~(\ref{zss1}, \ref{zrs1}).
However, the spins at the $t=0$ boundary now need not be completely aligned, and small domain walls can be created at the cost of a finite free energy per unit length.

Using the same graphical notation as in Fig.~\ref{fig:ZoverZ}(a), with an additional color, black, representing the ``free'' boundary condition $f$, and dashed blue line representing the finite strength boundary magnetic field $h_x$ in the ``$+$'' direction at $t=0$, we represent $\overline{\chi} = \frac{Z_{\rho\sigma}}{Z_{\sigma\sigma}}$ again with partition functions of appropriate boundary conditions in Fig.~\ref{fig:ZoverZ}(b).
First consider a case where the boundary magnetic field $h_x = -\frac{1}{2} \ln \lz \tr (\rho_x \cdot \sigma_x) \rz$ is uniform and independent of $x$.
This would be the case when, say, $\rho = \frac{1}{2^L} \mathbb{1}$ and $\sigma = (\ket{0}\bra{0})^{\otimes L}$.
Let the free energy cost of a domain wall with unit length be $\delta F$ and define the fugacity to be $y = e^{-\delta F}$, we have (compare Fig.~\ref{fig:ZoverZ}(b)) 
\begin{align}
\label{eq:ZoverZ_noencode_uniform_field}
    \overline{\chi} =&\ \frac{Z_{\rho\sigma}}{Z_{\sigma\sigma}} \nn
    =&\ \frac{1 + \binom{L}{1} e^{-2h} y +  \binom{L}{2} e^{-4h} y^2 + \ldots}{1 + \binom{L}{1} y +  \binom{L}{2} y^2 + \ldots} \nn
    \approx&\ \frac{(1+e^{-2h} y)^L}{(1+y)^L}.
\end{align}
Both $Z_{\rho\sigma}$ and $Z_{\sigma\sigma}$ numerator are now a series of terms, with the $i$-th leading term having $i$ domain walls each of unit length (neglecting their interactions).
Thus, $\overline{\chi}$ is exponentially suppressed by $L$ for any $h>0$, thus negligible throughout the phase diagram.

We can also generalize Eq.~\eqref{eq:ZoverZ_noencode_uniform_field} to the case where $\rho$ and $\sigma$ only differ on one site.
The partition functions are shown in Fig.~\ref{fig:ZoverZ}(c), where we obtain
\begin{align}
\label{eq:ZoverZ_noencode_local_field}
    \overline{\chi} \approx \frac{1+e^{-2h}}{2} + \frac{1-e^{-2h}}{2} \avg{s_{x\in \partial \mc{M}_0}}_{s_{x \in \partial \mc{M}_T} = +1}.
\end{align}
Here $\avg{s_{x\in \partial \mc{M}_0}}_{s_{x \in \partial \mc{M}_T} = +1}$ is the expectation value of a boundary spin.
Thus, we expect the following behavior of $\chi$ near the critical point:
\begin{align}
\label{eq:ZoverZ_noencode_local_field_result}
    \overline{\chi} \approx \begin{cases}
        |p-p_c|^\beta + \chi_0, &p < p_c \\
        \chi_0, &p > p_c
    \end{cases}.
\end{align}
Here, $\chi_0 \approx \frac{1+e^{-2h}}{2}$ is a nonuniversal constant between $0$ and $1$.
This expectation is confirmed by numerical results in Fig.~\ref{fig:ZoverZ}(d).

In this case, we do not expect a crossing as in Fig.~\ref{fig:chi_clifford_rho_stab_sigma_stab}, but instead a collapse of the curves for different system sizes $L$ (without rescaling the axes).
In experiments, a collapse is likely harder to detect than a crossing, for it will be more susceptible to noise for a given system size; that is, the collapse will immediately disappaear for any rate of noise.
For this reason, we have chosen to focus on the circuit with an encoding stage throughout the paper.

Moreover, for the purpose of observing MIPT, including the encoding stage should only introduce minor experimental overhead.
For example, noise in the encoding stage $t \in [0, t_{\rm encoding}]$ would not affect the signal for MIPT in any important way as its effect can be accounted for by a different choice of $\rho$, which is not essential (see discussions in Sec.~\ref{sec:map_chi_spin}); only noise in the circuit bulk $t \in [t_{\rm encoding}, t_{\rm encoding}+t_{\rm bulk}]$ is important (their effects shown Fig.~\ref{fig:chi_clifford_rho_stab_sigma_stab}(b,c)).

\subsection{Higher order cross entropies and the Kullback-Leibler divergence}

%The bulk cross entropy can be generalized to a generic index $Q$, 
Here we compute another measure of the difference between the two probability distributions $p^{\rho}_{\bs{m}}$ and $p^{\sigma}_{\bs{m}}$, the Kullback-Leibler divergence,
\begin{equation}
    D_{\mathrm{KL}}(p^{\rho}|p^{\sigma}) = \sum_{\bs{m}} (p^{\rho}_{\bs{m}} \log p^{\rho}_{\bs{m}} - p^{\rho}_{\bs{m}} \log p^{\sigma}_{\bs{m}}).
\end{equation}
This can be computed by the replica trick, where we introduce an integer replica index $Q$ and the higher-order cross entropies 
\begin{equation}
\label{eq:chiQ_app}
    \chi_Q = \mathbb{E}_C \frac{\sum_{\bs{m}} p^{\rho}_{\bs{m}}(p^{\sigma}_{\bs{m}})^{Q-1}}{\sum_{\bs{m}} (p^{\rho}_{\bs{m}})^Q}.
\end{equation}
When $Q=2$, this reduces to the linear cross entropy Eq.~\eqref{eq:chi_C} with the roles of $\rho$ and $\sigma$ exchanged. %For an integer $Q\geq 2$, this can be efficiently sampled using the hybrid quantum-classical algorithm in the main text. 
Thus, this quantity cannot be sampled using the method presented in the main text.

In order to understand the higher-order cross entropies in terms of the stat-mech model, we again resort to the annealed average
\begin{equation}
\label{eq:chibarQ_app}
    \overline{\chi}_Q =  \frac{\mathbb{E}_C\sum_{\bs{m}} p^{\rho}_{\bs{m}}(p^{\sigma}_{\bs{m}})^{Q-1}}{\mathbb{E}_C\sum_{\bs{m}} (p^{\rho}_{\bs{m}})^Q}
\end{equation}
Although the two quantities $\chi_{Q}$ and $\overline{\chi}_{Q}$ are different in general, they become the same quantity in the limit $Q\rightarrow 1$. More precisely, at $Q=1$, $\chi_Q = \overline{\chi}_Q =1$ and
\begin{equation}
    \left. \frac{d\chi_Q}{dQ}\right|_{Q=1} = \left. \frac{d\overline{\chi}_Q}{dQ}\right|_{Q=1} = - \mathbb{E}_C D_{\text{KL}}(p^{\rho}|p^{\sigma}).
\end{equation}
In order to see this, we expand Eq.~\eqref{eq:chiQ_app} to first order in $Q-1$,
\begin{widetext}
\begin{align}
    \chi_Q &\ = \mathbb{E}_C \frac{1 + (Q-1)\sum_{\bs{m}} p^{\rho}_{\bs{m}} \log p^{\sigma}_{\bs{m}}+ O((Q-1)^2)}{1 + (Q-1)\sum_{\bs{m}} p^{\rho}_{\bs{m}} \log p^{\rho}_{\bs{m}}+ O((Q-1)^2)} \nn
    &\ = 1+ (Q-1) \mathbb{E}_C \sum_{\bs{m}} (p^{\rho}_{\bs{m}} \log p^{\sigma}_{\bs{m}}-p^{\rho}_{\bs{m}} \log p^{\rho}_{\bs{m}})+ O((Q-1))^2 \nn
    &\ = 1 - (Q-1) \mathbb{E}_C D_{\text{KL}}(p^{\rho}|p^{\sigma}) + O((Q-1)^2),
\end{align}
\end{widetext}
%where $D_{\text{KL}}(p^{\sigma}|p^{\rho}) = (\sum_{\bs{m}} p^{\sigma}_{\bs{m}} \log p^{\sigma}_{\bs{m}}- p^{\rho}_{\bs{m}} \log p^{\sigma}_{\bs{m}})$ is the classical Kullback-Leibler divergence of the two probability distributions.
Expanding Eq.~\eqref{eq:chibarQ_app} to the first order, we obtain the same expression. 
%Thus 
%\begin{equation}
%    \left. \frac{d\chi_Q}{dQ}\right|_{Q=1} = \left. \frac{d\overline{\chi}_Q}{dQ}\right|_{Q=1} = - \mathbb{E}_C D_{\text{KL}}(p^{\sigma}|p^{\rho}).
%\end{equation}

Next, we compute $\overline{\chi}_Q$ using the stat mech model assuming Haar random gates in the circuit.
Notice that this quantity can be infinite in general for Clifford circuits.
The spins on the honeycomb lattice take on $Q!$ different values labelled by group elements of $S_Q$, with three-body ferromagnetic interactions on each downward-pointing triangles \cite{andreas2019hybrid, choi2019spin}. 
The model has an $S_Q \times S_Q$ symmetry, which is spontaneously broken in the volume law phase.
Assuming Eq.~\eqref{eq:rho_sigma_product_states}, and repeating the derivation on the boundary conditions that leads to Eq.~\eqref{eq:chi_ZoverZ}, we obtain
\begin{equation}
    \overline{\chi}_Q = \frac{Z_{Q,\rho\sigma}}{Z_{Q,\rho\rho}},
\end{equation}
where 
\begin{eqnarray}
 Z_{Q,\rho\sigma} &=& \sum_{g\in S_{Q-1}} Z_{Q,ge} \\
 Z_{Q,\rho\rho} &=& \sum_{g\in S_{Q}} Z_{Q,ge},
\end{eqnarray}
and $Z_{Q,ge}$ is the partition function of the spin model with fixed boundary condition $g \in S_Q$ on the bottom boundary and fixed boundary condition $e$ (identity permutation) on the top boundary, $S_{Q-1}$ is the subgroup of $S_Q$ of permutations that keeps the first element invariant.
In the volume law phase $(p<p_c)$, domain walls have finite tension, thus $Z_{Q,ge}/Z_{Q,ee}=O(e^{-L})$ for every $g\neq e$. As a result $Z_{Q,\rho\sigma}\approx Z_{Q,\rho\rho}\approx Z_{Q,ee}$ and $\bar{\chi}_Q = 1 + O(e^{-L})$. In the area law phase the partition functions with different boundary conditions are on the same order and $\bar{\chi}_Q$ is an order one number that depends on $Q$.
These results are %identical
completely analogous
with the case of $Q=2$ as they involve similar arguments.

At the critical point $(p = p_c)$, assuming periodic boundary conditions in the spatial direction, $Z_{ge}$ is a partition function of the CFT on a finite cylinder with width $L$ and length $T$. The partition function can be written in two equivalent forms \cite{cardy2004bcft}
\begin{eqnarray}
\label{eq:expansion_Z_ge_longtime_Cardy_states}
    Z_{Q,ge} & = & \sum_{\alpha} \langle \alpha|g\rangle \langle e|\alpha\rangle e^{-2\pi (\Delta_{Q,\alpha}-c_Q/12)T/L}\\
             & = & \sum_{\beta} N^{ge}_{\beta} e^{-\pi (h_{Q,\beta}-c_Q/24)L/T}.
\end{eqnarray}
In the first expression, $\alpha$ runs over \textit{bulk} operators, $|g\rangle$ and $\langle e|$ are Cardy states corresponding to the two fixed boundary conditions, $\Delta_{Q,\alpha}$ is the bulk scaling dimension, $c_Q$ is the central charge. In the second expression, $\beta$ runs over \textit{boundary} operators, $h_{Q,\beta}$ is the scaling dimension of the boundary operator, $N^{ge}_{\beta}$ is the multiplicity of the boundary condition changing operator from boundary condition $g$ to boundary condition $e$. The first expression is useful when $T\gg L$, then we only keep the ground state $\alpha=\mathbb{1}$ in the sum. The second expression is useful when $T\ll L$, then we only keep the leading boundary condition changing operator in the sum. Thus,
\begin{eqnarray}
    Z_{Q,ge} =
    \begin{cases}
        e^{s_{Q,e}+s_{Q,g}} e^{-\pi c_Q T/(6L)}~~(T\gg L) \\
        e^{-\pi (h_{Q,ge}-c_Q/24)L/T} ~~(T\ll L)
    \end{cases},
\end{eqnarray}
where $s_{Q,g}\equiv \log (\langle \mathbb{1}|g\rangle)$ is known as the Affleck-Ludwig boundary entropy~\cite{AffleckLudwig1991boundaryentropy}, 
% is boundary entropy of the boundary condition $i$, 
$h_{Q,ge}$ is the scaling dimension of the lowest boundary condition changing operator from $g$ to $e$. Thus, at replica index $Q\geq 2$, we obtain 
\begin{equation}
\label{eq:chi_Q_long_time}
    \overline{\chi}_Q = \frac{\sum_{g\in S_{Q-1}} e^{s_{Q,g}}}{\sum_{g\in S_{Q}} e^{s_{Q,g}}} ~~ (T\gg L)
\end{equation}
and 
\begin{equation}
\label{eq:chi_Q_short_time}
    \overline{\chi}_Q = \frac{\sum_{g\in S_{Q-1}} e^{-\pi h_{Q,ge}L/T}}{\sum_{g\in S_{Q}}e^{-\pi h_{Q,ge}L/T} } ~~ (T\ll L)
\end{equation}

We focus on the long time limit $T \gg L$, starting with Eq.~\eqref{eq:chi_Q_long_time}.
Since we are looking at the critical point, the symmetry is not spontaneously broken, and the ground state $\ket{\mathbb{1}}$ is invariant under the action of $S_Q$.
Thus, $s_{Q,g}$ for all fixed boundary conditions $g$ are the same, and we have %due to the $S_Q$ symmetry, thus
\begin{equation}
    \overline{\chi}_Q = \frac{1}{Q} ~~ (T\gg L)
\end{equation}
% at the critical point.
% \YZ{The same result holds in the area law phase using similar argument}. Thus
The situation is similar in the area law phase $p > p_c$, where an expansion similar to Eq.~\eqref{eq:expansion_Z_ge_longtime_Cardy_states} can be written down from the transfer matrix of the spin model.
The spectrum here will become gapped, but in this limit only the ground state contributes.
Moreover, the ground state still preserves the symmetry of the model.
Summaring, we have
\begin{align}
    \overline{\chi}_Q = \begin{cases}
        1, & T\gg L, p < p_c \\
        1/Q, & T\gg L, p \geq p_c
    \end{cases},
\end{align}
% ignoring exponentially small terms. 
neglecting terms that are exponentially small in $L$.
% \YZ{This looks very strong. Not sure if this is correct. Can we check it numerically?}
Taking the $Q\to 1$ limit we obtain
\begin{equation}
\label{eq:KL}
    \mathbb{E}_C D_{\text{KL}}(p^{\rho}|p^{\sigma}) = \begin{cases}
        0, & T\gg L, p < p_c \\
        1, & T\gg L, p \geq p_c
    \end{cases}.
\end{equation}

Another closely related quantity is
\begin{equation}
\label{eq:chiQ_app_2}
    \chi'_Q = \mathbb{E}_C \frac{\sum_{\bs{m}} p^{\rho}_{\bs{m}}(p^{\sigma}_{\bs{m}})^{Q-1}}{\sum_{\bs{m}} (p^{\sigma}_{\bs{m}})^Q}.
\end{equation}
This is reduced to Eq.~\eqref{eq:chi_C} when $Q=2$. For generic integer $Q\geq 2$, it can be sampled efficiently using the hybrid quantum-classical algorithm in the main text,
\begin{equation}
    \chi'_Q = \left\langle\frac{(p^{\sigma}_{\bs{m}})^{Q-1}}{\sum_{\bs{m}} (p^{\sigma}_{\bs{m}})^Q} \right\rangle_{\rho}.
\end{equation}
% given that the circuit is Clifford and $\sigma$ is a stabilizer state.
As in $Q=2$, it is expected that the sampling error decays as $M^{-1/2}$, where $M$ is the sample size, since we are averging over random numbers with $O(1)$ amplitude. In order to interpret this quantity in the stat mech model, we consider the annealed average
\begin{equation}
\label{eq:chibarQ_app_2}
    \overline{\chi}'_Q =  \frac{\mathbb{E}_C\sum_{\bs{m}} p^{\rho}_{\bs{m}}(p^{\sigma}_{\bs{m}})^{Q-1}}{\mathbb{E}_C\sum_{\bs{m}} (p^{\sigma}_{\bs{m}})^Q}.
\end{equation}
The quenched and annealed averages again coincide in the replica limit,
\begin{equation}
    \frac{d\chi'}{dQ}= \frac{d\bar{\chi}'}{dQ} = \mathbb{E}_C \sum_{\bs{m}}(p^{\rho}_{\bs{m}}-p^{\sigma}_{\bs{m}}) \log p^{\sigma}_{\bs{m}}.
\end{equation}
This also measures the difference between the two probablity distributions, although it is not the KL divergence anymore. As long as $\rho$ and $\sigma$ are pure product states that are not identical, Eq.~\eqref{eq:chibarQ_app_2} and Eq.~\eqref{eq:chibarQ_app} are mapped to the same quantity in terms of the stat mech model. The argument that leads to Eq.~\eqref{eq:KL} goes through, and we obtain
\begin{align}
    \overline{\chi}'_Q = \begin{cases}
        1, & T\gg L, p < p_c \\
        1/Q, & T\gg L, p \geq p_c
    \end{cases},
\end{align}
% ignoring exponentially small terms. 
neglecting terms that are exponentially small in $L$.
% \YZ{This looks very strong. Not sure if this is correct. Can we check it numerically?}
Taking the $Q\to 1$ limit we obtain
\begin{equation}
\label{eq:KL}
    \mathbb{E}_C \sum_{\bs{m}}(p^{\sigma}_{\bs{m}}-p^{\rho}_{\bs{m}}) \log p^{\sigma}_{\bs{m}} = \begin{cases}
        0, & T\gg L, p < p_c \\
        1, & T\gg L, p \geq p_c
    \end{cases}.
\end{equation}
\subsection{Linear cross entropy with ancilla}
In this Appendix we investigate how coupling to an ancilla may affect the signature of the transition. Suppose we extend the system, $S$, with a system of ancilla qudits $W$, so that the initial state is $\rho_{SW}$ or $\sigma_{SW}$.
%We will not make specific assumptions on these states for now, but in general the system $S$ and the ancilla $W$ will be entangled.
In particular, we take $|W| = |S| = L$, and associate one ancilla qudit to each system qudit, collectively denoted as $SW,x$ on site $x$.
The evolution acts on system $S$ as before, while the ancillae $W$ are idlers, i.e. no evolution occurs. We denote the unnormalized output state when reduced to $W$ by $\rho_{\bs m W}= \tr_S (C_{\bs m}\rho_{SW}C^\dag_{\bs m})$, and we similarly define $\sigma_{\bs m W}$.
We consider the quantity
\be \chi'=\frac{\sum_{\bs m} \tr_W \rho_{\bs m W}\sigma_{\bs m W}}{\sum_{\bs m}\tr_W\sigma^2_{\bs m W}}\ee
Differently from Eq.~\eqref{eq:chi_C}, which is a cross entropy between classical probability distributions, here $\chi'$ is a ``quantum'' cross entropy.

We shall consider two particular ways of coupling system and ancilla ---one quantum and one classical. Each way of coupling will lead to a particular dependence of $\chi'$ on $p$ which in turn may be used to diagnose the transition.
%and can be probed provided that joint, entangled measurements can be made on the quantum state $\rho_{\bs m W} \otimes \sigma_{\bs m W}$ supported on the ancillae.
% Provided such measurements can be made, it is possible one can perform fewer runs $M$ to estimate $\chi'$, as such measurements will reveal more information about the initial state.
% However,  that $\chi'$ has smaller relative fluctuations compared to $\chi$, i.e. $\varepsilon_{\chi'}/\chi' <\varepsilon_{\chi}/\chi$.
% The possible advantage would be that, at the cost of preparing the state of a bigger system $\rho_{SW}$, one can perform fewer runs $M$ to estimate $\chi'$.
%
%Let us consider a concrete example where $|W| = |S| = L$, and $\sigma_{SW}=\prod_{x=1}^L\sigma_{SW,x}$ as well as $\rho_{SW}=\prod_{x=1}^L\rho_{SW,x}$.
We will always take initial product states $\rho_{SW}=\prod_x \rho_{SW,x}$ and $\sigma_{SW}=\prod_x \sigma_{SW,x}$; moreover, for simplicity we shall asssume that the ancilla is decoupled from the system in the $\rho$ state $\rho_{SW}=\rho_S\otimes \rho_W$.
We will study two forms of $\sigma_{SW}$: 1) EPR state for the ancilla-system on each site each site $x$, $\sigma_{SW, x}=|I_x\rangle\langle I_x|$ with $\ket{I_ x}=\frac 1{\sqrt d}\sum_i|i_{S,x}i_{W,x}\rangle$, and 2) classically correlated state $\sigma_{SW, x}=\frac 1d\sum_i|i_{S,x}i_{W,x}\rangle\langle i_{S,x}i_{W,x}|$.

In the EPR pair case, we have (suppressing the $x$-dependence $\rho_{SW,x}\to\rho_{SW}$ and $\sigma_{SW,x}\to\sigma_{SW}$)
\begin{widetext}
\begin{align}
    Z_{\rho \sigma} =&\ \sum_{\{s_i = \pm 1\}} \prod_{\langle i, j, k \rangle \in \triangledown} J_p(s_i, s_j; s_k) 
    \cdot \prod_{x \in \partial \mc{M}_T} \delta_{s_x=+1}
    \cdot \prod_{x \in \partial \mc{M}_0} \tr_W\(\delta_{s_x=+1} (\tr_S \rho_{SW})(\tr_S \sigma_{SW}) + \delta_{s_x=-1} \tr_S (\rho_{SW} \cdot \sigma_{SW}) \) \nn
    =&\ \sum_{\{s_i = \pm 1\}} \prod_{\langle i, j, k \rangle \in \triangledown} J_p(s_i, s_j; s_k) 
    \cdot \prod_{x \in \partial \mc{M}_T} \delta_{s_x=+1}
    \cdot \prod_{x \in \partial \mc{M}_0}\left( \delta_{s_x=+1}\cdot\frac 1d +\delta_{s_x=-1}\cdot \frac 1d \tr(\rho_S\rho_W)\right)=d^{-L} Z_{++}\\
    Z_{\sigma \sigma} =&\ \sum_{\{s_i = \pm 1\}} \prod_{\langle i, j, k \rangle \in \triangledown} J_p(s_i, s_j; s_k) 
    \cdot \prod_{x \in \partial \mc{M}_T} \delta_{s_x=+1}
    \cdot \prod_{x \in \partial \mc{M}_0} \tr_W\(\delta_{s_x=+1} (\tr_S \sigma_{SW})(\tr_S \sigma_{SW}) + \delta_{s_x=-1} \tr_S (\sigma_{SW} \cdot \sigma_{SW}) \) \nn
    =&\ \sum_{\{s_i = \pm 1\}} \prod_{\langle i, j, k \rangle \in \triangledown} J_p(s_i, s_j; s_k) 
    \cdot \prod_{x \in \partial \mc{M}_T} \delta_{s_x=+1}
    \cdot \prod_{x \in \partial \mc{M}_0} \left(\delta_{s_x=+1}\cdot\frac 1d+\delta_{s_x=-1}\right)=d^{-L}Z_{++}+Z_{-+}\,.
\end{align}
\end{widetext}
Where in the last step of each of the above equations we assumed $\tr(\rho_S\rho_W)<1$. We see that 
\be\ovl\chi=\frac{Z_{\rho\sigma}}{Z_{\sigma\sigma}}=\frac{Z_{++}}{Z_{++}+d^L Z_{-+}}\,.\ee
%At $p=0$, $d^LZ_{-+}=d^Ld^{-L}=Z_{++}$, so $\bar\chi=1/2$, and we find
%\be \bar\chi=\begin{cases}1,&p=0\\
%d^{-O(L)},&p>0\,.\end{cases}\ee
The tension of domain walls in the Ising model (see Eq.~\eqref{fig:ZoverZ}) decreases as $p$ becomes finite, therefore $Z_{-+}\sim e^{- a L\log d}$ with $a<1$. Therefore, as soon as $p>0$, $\bar\chi$ becomes exponentially suppressed in system size $L$ thus destroying the signature of the transition. This is because at the final time we have access to sufficient information about the initial quantum state of the system due to the highly-entangled ancilla-system coupling.

For the classically correlated state $\sigma_{SW}=\frac 1d\sum_i|i_Si_W\rangle\langle i_Si_W|$, we have
\begin{widetext}
\begin{align}
    Z_{\rho \sigma} =&\ \sum_{\{s_i = \pm 1\}} \prod_{\langle i, j, k \rangle \in \triangledown} J_p(s_i, s_j; s_k) 
    \cdot \prod_{x \in \partial \mc{M}_T} \delta_{s_x=+1}
    \cdot \prod_{x \in \partial \mc{M}_0} \tr_W\(\delta_{s_x=+1} (\tr_S \rho_{SW})(\tr_S \sigma_{SW}) + \delta_{s_x=-1} \tr_S (\rho_{SW} \cdot \sigma_{SW}) \) \nn
    =&\ \sum_{\{s_i = \pm 1\}} \prod_{\langle i, j, k \rangle \in \triangledown} J_p(s_i, s_j; s_k) 
    \cdot \prod_{x \in \partial \mc{M}_T} \delta_{s_x=+1}
    \cdot \prod_{x \in \partial \mc{M}_0}\left( \delta_{s_x=+1}\cdot\frac 1d +\delta_{s_x=-1}\cdot \frac 1d \tr(\tilde \rho_S\tilde \rho_W)\right)=d^{-L} Z_{++}\\
    Z_{\sigma \sigma} =&\ \sum_{\{s_i = \pm 1\}} \prod_{\langle i, j, k \rangle \in \triangledown} J_p(s_i, s_j; s_k) 
    \cdot \prod_{x \in \partial \mc{M}_T} \delta_{s_x=+1}
    \cdot \prod_{x \in \partial \mc{M}_0} \tr_W\(\delta_{s_x=+1} (\tr_S \sigma_{SW})(\tr_S \sigma_{SW}) + \delta_{s_x=-1} \tr_S (\sigma_{SW} \cdot \sigma_{SW}) \) \nn
    =&\ \sum_{\{s_i = \pm 1\}} \prod_{\langle i, j, k \rangle \in \triangledown} J_p(s_i, s_j; s_k) 
    \cdot \prod_{x \in \partial \mc{M}_T} \delta_{s_x=+1}
    \cdot \prod_{x \in \partial \mc{M}_0} \left(\delta_{s_x=+1}\cdot\frac 1d+\delta_{s_x=-1}\cdot \frac 1d\right)=d^{-L}Z_{++}+d^{-L}Z_{-+}\,,
\end{align}
\end{widetext}
where $\tilde\rho=\sum_i\langle i|\rho|i\rangle|i\rangle\langle i|$ denotes a dephased state, and we assumed $\tr(\tilde\rho_S\tilde\rho_W)<1$. We now have
\be\ovl\chi=\frac{Z_{\rho\sigma}}{Z_{\sigma\sigma}}=\frac{Z_{++}}{Z_{++}+ Z_{-+}}\ee
and we find the same behavior as in the case without ancilla. We conclude that a classically correlated ancilla does not seem to qualitatively alter the signature of the transition. This is expected, as at the final time we only have access to classical information about the initial state.

\section{Numerical algorithm for $\chi_C$ in Clifford circuits \label{sec:clifford_alg_chi}}

We first recall a ``purified'' representation of the hybrid circuit.
As pointed out in Refs.~\cite{gullans2019purification, choi2019spin}, the dynamics of the hybrid circuit can be purified by introducing one ``register'' qubit for each single site measurement.
In particular, each measurement can be replaced by a controlled-NOT (CNOT) gate from the measured qubit to the register, followed by a dephasing channel on the register.
\footnote{To see this, it is sufficient to consider the case of one qubit and one register.
Initially, let the qubit be in the state $\ket{\psi} = \alpha \ket{0} + \beta \ket{1}$, and the register be in the state $\ket{0}$, so the joint state is
\begin{align}
    \rho_{QR} =  \ket{\psi}\bra{\psi}_Q
    \otimes
    \ket{0}\bra{0}_R.
\end{align}
After the CNOT cate, we have
\begin{align}
    \rho_{QR}^\p =&\  \mathrm{CNOT}_{Q\to R} \cdot \rho_{QR} \cdot \mathrm{CNOT}_{Q\to R} \nn
    =&\ 
    \(\alpha\ket{00} + \beta\ket{11} \)
    \(\alpha^\ast\bra{00} + \beta^\ast\bra{11} \)_{QR}.
\end{align}
Under the dephasing channel on $R$, 
\begin{align}
    \rho_{QR}^{\p\p}
    =&\ 
    \frac{1}{2}\(\rho_{QR}^{\p} +  Z_R \rho_{QR}^{\p} Z_R\)\nn
    =&\ 
    |\alpha|^2 \ket{0}\bra{0}_Q \otimes \ket{0}\bra{0}_R
    +
    |\beta|^2 \ket{1}\bra{1}_Q \otimes \ket{1}\bra{1}_R \nn
    =&\
    \(P_0 \ket{\psi}\bra{\psi} P_0\)_Q \otimes \ket{0}\bra{0}_R
    +
    \(P_1 \ket{\psi}\bra{\psi} P_1\)_Q \otimes \ket{1}\bra{1}_R.
\end{align}
The result $\rho_{QR}^{\p\p}$ is a mixture of different trajectories, with the measurement outcome stored in $R$.
Generalization to many qubits and many registers can be carried out in a similar fashion.}
With these, at the end of the time evolution we have the following joint state on physical qubits $Q$ and register qubits $R$,
\begin{equation}
    \rho_{QR} = \sum_{\bs{m}} C_{\bs{m}} \rho C_{\bs{m}}^\dg \otimes \ket{\bs{m}}\bra{\bs{m}}_R.
\end{equation}
And similarly for the initial state $\sigma$,
\begin{align}
    \sigma_{QR} = \sum_{\bs{m}} C_{\bs{m}} \sigma C_{\bs{m}}^\dg \otimes \ket{\bs{m}}\bra{\bs{m}}_R.
\end{align}
The cross entropy will then have the following representation
\begin{align}
\label{eq:chi_C_stabilizer_register}
    \chi_C 
    = \frac{
        \sum_{\bs{m}} p_{\bs{m}}^\rho p_{\bs{m}}^\sigma
    }
    {
        \sum_{\bs{m}} \(p_{\bs{m}}^\sigma\)^2
    }
    = \frac{\tr \rho_R \sigma_R}{\tr \sigma_R^2},
\end{align}
where $\rho_R$ is the reduced state of $\rho_{QR}$ on $R$, and similarly for $\sigma_R$.

We now focus on the case where $\rho$ and $\sigma$ are both stabilizer states and the circuit is a Clifford circuit, so that $\rho_{QR}, \sigma_{QR}, \rho_R, \sigma_R$ are all stabilizer states.
Moreover, we will choose the state $\rho$ to be obtainable from $\sigma$ via erasure and dephasing channels.
Equivalently, we choose states $\rho$ and $\sigma$ such that the stabilizer group $\mc{S}_\rho$ is a subgroup of $\mc{S}_\sigma$.
Whenever this condition is satisfied for the initial state, it follows that $\mc{S}_{\rho_{QR}} \subseteq \mc{S}_{\sigma_{QR}}$ and $\mc{S}_{\rho_{R}} \subseteq \mc{S}_{\sigma_{R}}$ at any point of the purified circuit evolution.
With this property, Eq.~\eqref{eq:chi_C_stabilizer_register} can be greatly simplified.
We have
\begin{align}
    \rho_R = \frac{1}{2^{|R|}} \sum_{g \in \mc{S}_{\rho_R}} g, \quad
    \sigma_R
    = \frac{1}{2^{|R|}} \sum_{h \in \mc{S}_{\sigma_R}} h,
\end{align}
and
\begin{align}
\label{eq:tr_rho_R_sigma_R}
    \tr \rho_R \sigma_R =&\  \frac{1}{2^{2|R|}} \sum_{g \in \mc{S}_{\rho_R}} \sum_{h \in \mc{S}_{\sigma_R}} \tr gh 
    \nn
    =&\ \frac{1}{2^{2|R|}} \sum_{g \in \mc{S}_{\rho_R}} \sum_{h \in \mc{S}_{\rho_R}} \tr gh
    \nn 
    =&\ \tr \rho_R^2.
\end{align}
Here, we noticed that $\tr gh = 2^{|R|}  (\delta_{g, h} - \delta_{g, -h})$ for Pauli strings $g$ and $h$, and used $\mc{S}_{\rho_R} \subseteq \mc{S}_{\sigma_R}$.
Thus, the cross entropy is simply the ratio between the second Renyi purity of the probability distributions $\{p^\rho_\bs{m}\}$ and $\{p^\sigma_\bs{m}\}$,
\begin{align}
    \chi_C = \frac{\tr \rho_R^2}{\tr \sigma_R^2}
    =
    \frac{
        \sum_{\bs{m}} \(p_{\bs{m}}^\rho\)^2
    }
    {
        \sum_{\bs{m}} \(p_{\bs{m}}^\sigma\)^2
    }.
\end{align}
For Clifford circuit evolution, the second Renyi purity equals $2^{-N_{\rm rand}}$, where $N_{\rm rand}$ is the number of measurements (out of the total $N$) whose outcome is randomly $\pm 1$ (see footnote~\ref{fn:Nrand}).
This number $N_{\rm rand}$ can be obtained by running the circuit once for each initial state~\cite{aaronson2004chp}.
We have
\begin{align}
\label{eq:chi_C_Nrand}
    \chi_C = 2^{-N_{\rm rand}(C, \rho) + N_{\rm rand}(C, \sigma)}.
\end{align}

%----------------------------------
\begin{figure}[t]
    \centering
    \includegraphics[width=.45\textwidth]{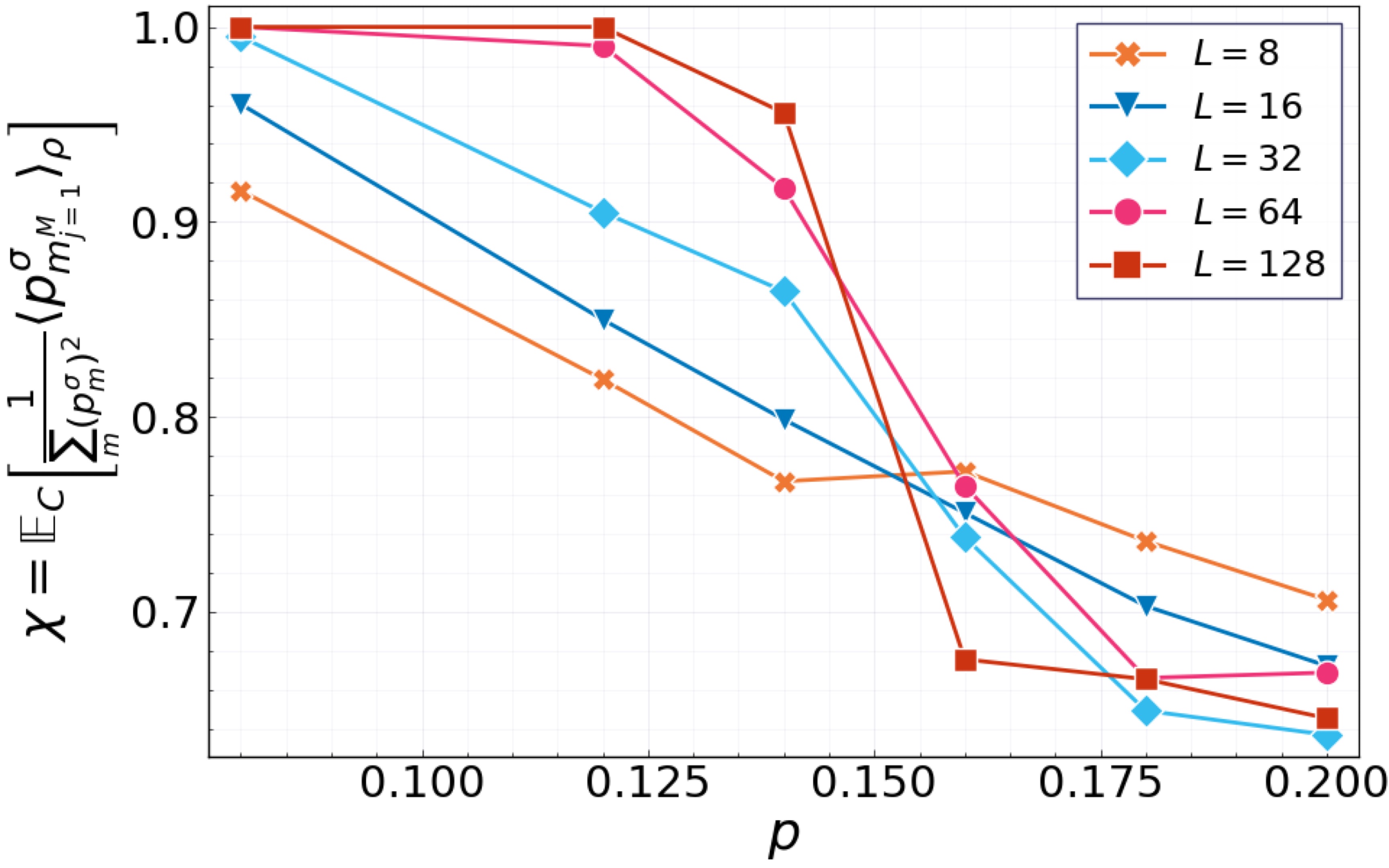}
    \caption{Numerical results of $\chi$ for initial states $\rho = (\ket{0}\bra{0})^{\otimes L/2} \otimes (\ket{+}\bra{+})^{\otimes L/2}$ and $\sigma =(\ket{0}\bra{0})^{\otimes L}$, following the procedure in Eq.~\eqref{eq:chi_C_clifford_sample}.
    Despite a different choice of initial state, the results are comparable to Fig.~\ref{fig:chi_clifford_rho_stab_sigma_stab}(a) and Fig.~\ref{fig:chi_clifford_rho_magic_sigma_stab}.
    Here the number of circuir realizations is 300, and for each circuit $M = 100$ runs of the circuit are taken.
    }
    \label{fig:clifford_numerics_hadamard}
\end{figure}
%----------------------------------

More generally, for initial states $\rho$ and $\sigma$ that may not satisfy the condition $\mc{S}_\rho \subseteq \mc{S}_\sigma$, Eq.~\eqref{eq:tr_rho_R_sigma_R} takes the form
\begin{align}
    \tr \rho_R \sigma_R
     =&\  \frac{1}{2^{2|R|}} \sum_{g \in \mc{S}_{\rho_R}} \sum_{h \in \mc{S}_{\sigma_R}} \tr gh
    = 
    \begin{cases}
        \frac{|\mc{S}_{\rho_R} \cap \mc{S}_{\sigma_R} |}{2^{|R|}}, & -1 \notin \mc{S}_{\rho_R} \cdot \mc{S}_{\sigma_R} \\
        0, & -1 \in \mc{S}_{\rho_R} \cdot \mc{S}_{\sigma_R}  \\
    \end{cases}.
\end{align}
Computing this without further simplifications can take time $O(L^3 \cdot T^3)$.
In practice it would be most convenient to carry out the sampling procedure outlined in Eq.~\eqref{eq:chi_C_clifford_sample},
\begin{align}
% \label{eq:chi_C_clifford_sample_rewrite}
    \chi_C = \lim_{M \to \infty} \avg{ \frac{p^\sigma_{\bs{m}_{j=1}^M}} {\sum_{\bs{m}} \(p_{\bs{m}}^\sigma\)^2}
    }_\rho. %\nonumber
    %= 
    % \frac{1}{\tr \sigma_R^2} \lim_{M \to \infty} \avg{p^\sigma_{\bs{m}_{j=1}^M} }_\rho.
\end{align}
which, as we have shown, converges in $\mathrm{poly}(1/\varepsilon)$ time.
That is, we run the $\rho$-circuit and obtain an ensemble of measurement histories $\{\bs{m}_{j} \}$, and take the average of their corresponding probabilities $p^\sigma_{\bs{m}_{j}}$ in the $\sigma$-circuit, divided by $\sum_{\bs{m}} \(p_{\bs{m}}^\sigma\)^2 = 2^{-N_{\rm rand}}$.
Each $p^\sigma_{\bs{m}_{j}} / \sum_{\bs{m}} \(p_{\bs{m}}^\sigma\)^2$ can be computed in polynomial time by running a $\sigma$-circuit in parallel.

To verify the validity of this method, we consider initial states $\rho = (\ket{0}\bra{0})^{\otimes L/2} \otimes (\ket{+}\bra{+})^{\otimes L/2}$ and $\sigma =(\ket{0}\bra{0})^{\otimes L}$.
Both are stabilizer states, but $\mc{S}_\rho \not\subseteq \mc{S}_\sigma$, and Eq.~\eqref{eq:chi_C_Nrand} does not apply.
We carry out the sampling procedure in Eq.~\eqref{eq:chi_C_clifford_sample}, and plot the results
in Fig.~\ref{fig:clifford_numerics_hadamard}, which we find comparable to Fig.~\ref{fig:chi_clifford_rho_stab_sigma_stab}(a) despite a more involved numerical calculation.
Thus, to estimate $\chi$ we have the freedom of choosing $\rho$, as consistent with the picture developed in  Sec.~\ref{sec:map_chi_spin}.

\section{Bitstring distribution in the output state \label{sec:porter_thomas}}

%----------------------------------------
\begin{figure*}
    \centering
    \includegraphics[width=.40\textwidth]{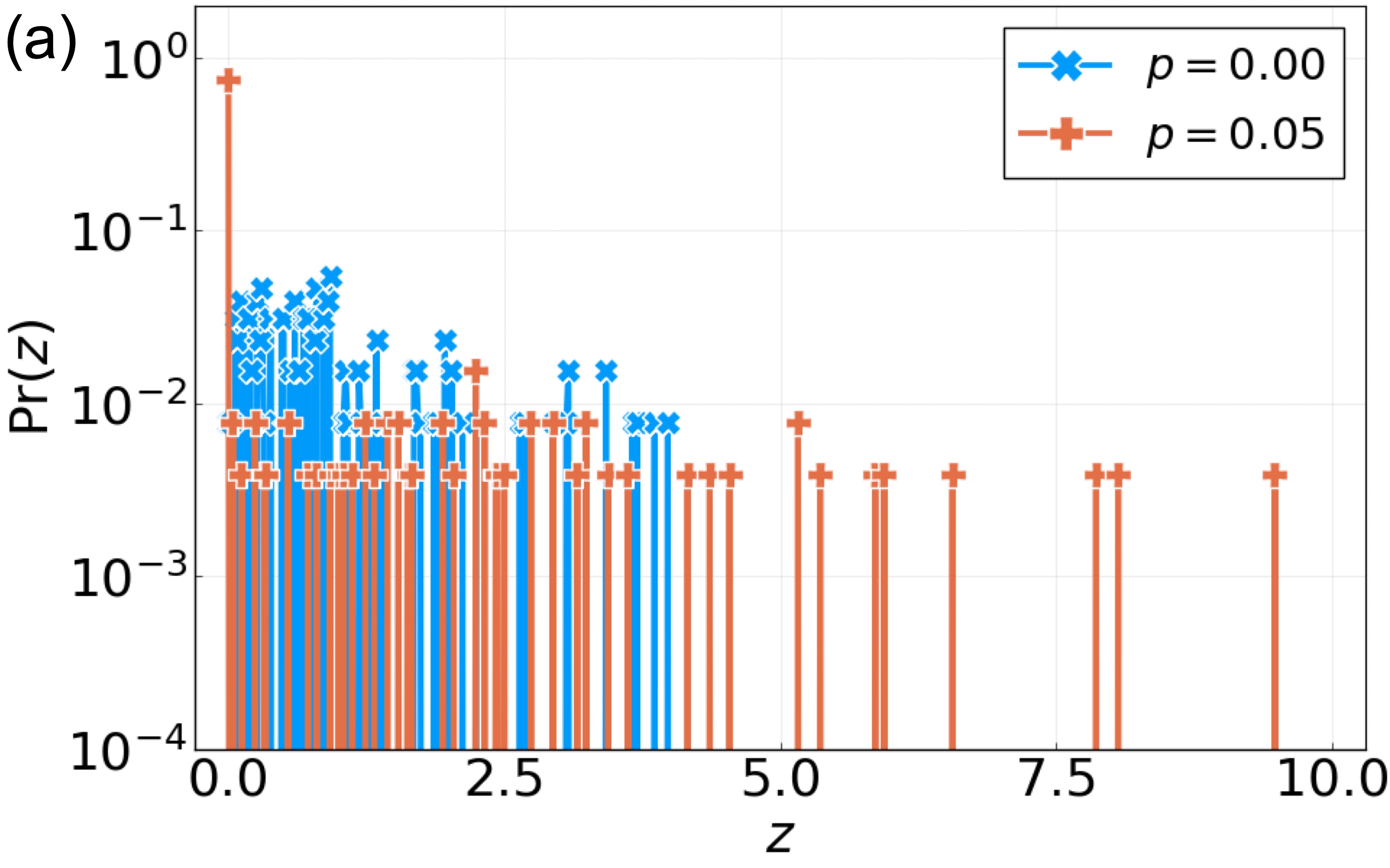}
    \quad \quad
    \includegraphics[width=.40\textwidth]{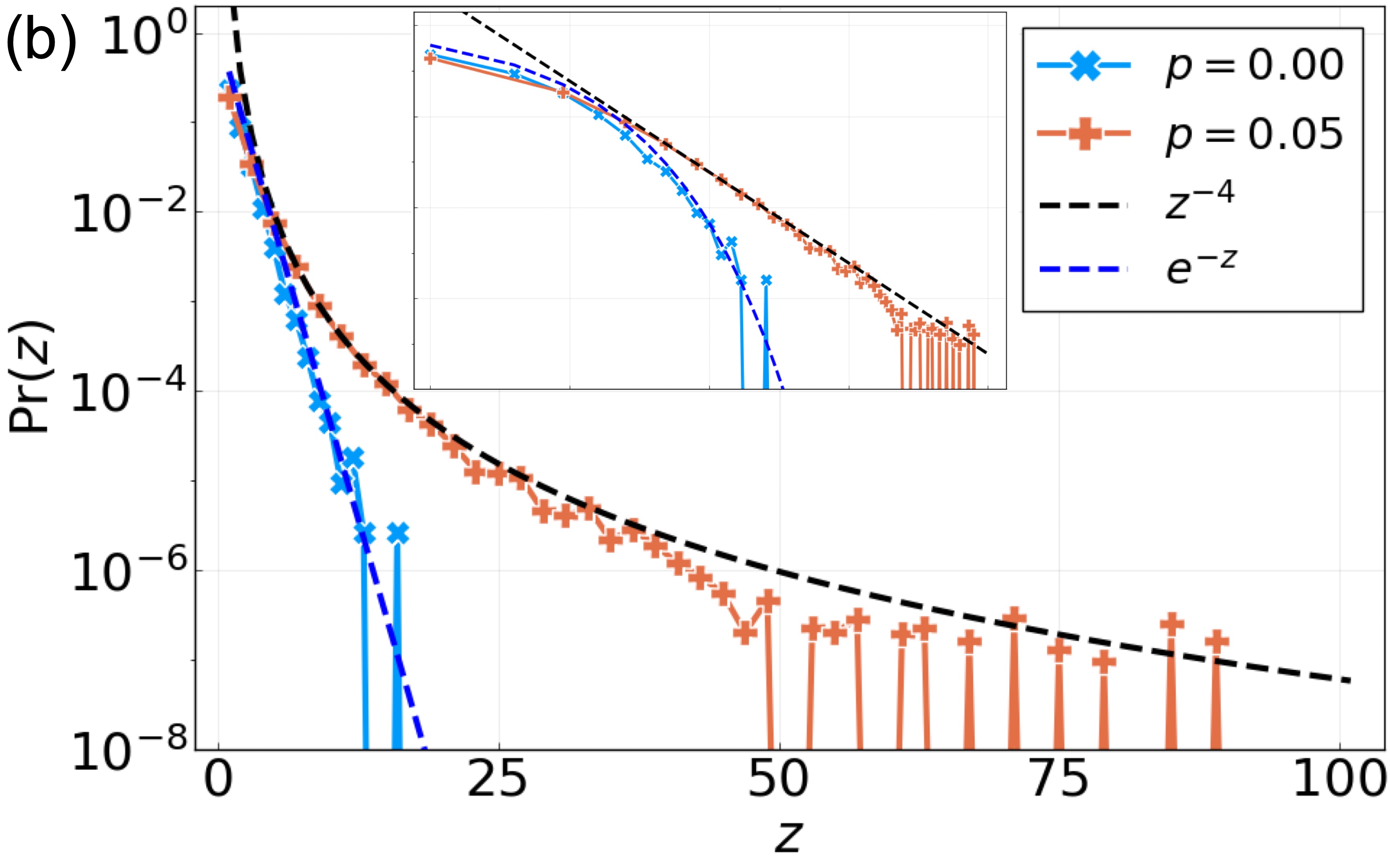}
    % ~
    % \includegraphics[width=.32\textwidth]{figs/F_XEB}
    % \includegraphics[width=.45\textwidth]{figs/F_diff_hist}
    \caption{(a)
    The bitstring distribution defined in Eq.~\eqref{eq:def_PrZ_typical}, for a typical instance of $C_{\bs{m}}$ with a generic (nonstabilizer) initial state $\rho$ in Eq.~\eqref{eq:rho_T_gate_half_L}.
    We see a broad distribution, in sharp contrast to the bitstring distribution from a stabilizer initial state in Eq.~\eqref{eq:def_PrZ_typical_stabilizer}.
    (b)
    The bitstring distribution defined in Eq.~\eqref{eq:def_Pr_z}, when the data in (a) is averaged over $C_{\bs{m}}$.
    The Porter-Thomas distribution $ \mathrm{Pr}(z) = e^{-z}$ is reproduced in the unitary limit $p = 0$, and a qualitatively different (powerlaw) distribution is observed for $p > 0$ (see Eq.~\eqref{eq:Pr_z_p>0}).
    % (c)
    % The circuit-averaged linear cross entropy $\chi_{\rm out}$ defined in Eq.~\eqref{eq:def_chi_out} between the output of a sampler $\psi$ and the target distribution $\mu(x; C_{\bs{m}}, \rho)$.
    % Here we compare two cases: when $\psi$ is the quantum-simulable $\rho$-circuit that outputs ${\rho}_{\bs{m}}$ (for which we compute $\mathbb{E}_{C_{\bs{m}}} \chi_{\rm out}(\rho_{\bs{m}})$), and when $\psi$ is the classically-simulable $\sigma$-circuit that outputs $\sigma_{\bs{m}}$ (for which we compute $\mathbb{E}_{C_{\bs{m}}} \chi_{\rm out}(\sigma_{\bs{m}})$).
    % They confirm the relation in Eq.~\eqref{eq:22} as derived from a mapping to stat mech model.
    % In the inset, we show that even without the circuit average, the difference between the two for a typical circuit $C_\bs{m}$, namely  $\Delta \chi_{\rm out} \coloneqq \chi_{\rm out}(\rho_{\bs{m}}) - \chi_{\rm out}(\sigma_{\bs{m}})$, is already large and positive.
    }
    \label{fig:PrZ}
\end{figure*}
%----------------------------------------

As we discussed in the main text, the linear cross entropy $\chi$ for the MIPT is most conveniently estimated numerically for Clifford circuits with a stabilizer initial state $\rho$, and can be extended to Clifford circuit with a non-stabilizer $\rho$ (and scaled up) given access to a quantum processor.
In either case $\chi$ admits the same interpretation in the stat mech language, and should contain the same universal data, e.g. the critical exponent $\nu$.
Thus, one natural question is whether considering a non-stabilizer initial state on a quantum processor reveals anything new about the physics surrounding the MIPT.

% Here we consider one such possibility, focusing on the volume law phase $p < p_c$.
As we have shown, in the volume law phase, $\chi = 1$ almost identically for sufficiently large $L$; and it follows that it is impossible -- in an information-theoretic sense -- to distinguish two different initial states from infrequent ($p<p_c$) bulk measurements.
The information about the initial state must therefore be contained in the output state of the circuit.

The difference  between the two initial states may be detected using various measures~\cite{zhou2019single, iaconis2020statecomplexity}.
Here we consider the probability distribution over bitstrings when each qubit of the output state of the $\rho$-circuit (namely $\rho_{\bs{m}} =  C_{\bs{m}} \rho C_{\bs{m}}^\dg$ in Eq.~\eqref{eq:def_rho_m}) is measured in the computational basis, where the input state $\rho$ is taken to be the one from Eq.~\eqref{eq:rho_T_gate_half_L} in the main text.
For a fixed bitstring $x \in \{0,1\}^L$, the probability for this outcome to occur in the output state of $C_{\bs{m}}$ is 
\begin{align}
    \mu
    (x; C_\bs{m}, \rho) = \bra{x} \overline{\rho}_{\bs{m}} \ket{x},
\end{align}
where $\overline{\rho}_{\bs{m}} =  \rho_{\bs{m}} / \tr \rho_{\bs{m}}$ is the normalized output state.
% \st{Define $z \coloneqq \mu (x; C_\bs{m}, \rho) \cdot D$, where $D = 2^L$ is the dimension of the $L$-qubit Hilbert space.}
In Fig.~\ref{fig:PrZ}(a)
we plot the fraction of bistrings with probability  $\mu = z/D$ in a typical instance of $C_{\bs{m}}$, where $z$ is a random variable and $D = 2^L$ is the dimension of the $L$-qubit Hilbert space,
\begin{align}
\label{eq:def_PrZ_typical}
    \mathrm{Pr}(z; C_{\bs{m}}, \rho) = \frac{1}{D} \sum_{x \in \{0,1\}^L} \delta(z - \mu (x; C_\bs{m}, \rho) \cdot D).
\end{align}
As we can see, in a typical circuit at $p > 0$ the output distribution is already notably broader than at $p=0$.

On the other hand, for the output of the $\sigma$-circuit, namely $\overline{\sigma}_{\bs{m}} = C_{\bs{m}} \sigma C_{\bs{m}}^\dg / \tr C_{\bs{m}} \sigma C_{\bs{m}}^\dg$ where $\sigma$ is a stabilizer state, the distribution function $\mathrm{Pr}(z; C_{\bs{m}}, \sigma)$ is much simpler:
% .In this case, there are $2^{+n}$ (where $n \le L$ is an integer) bitstrings that all have probability equal to $2^{-n}$; and all other bitstrings have zero probability.
% Thus,
\begin{align}
\label{eq:def_PrZ_typical_stabilizer}
\mathrm{Pr}(z; C_{\bs{m}}, \sigma) = \( 1-\frac{1}{2^{L-n}} \) \delta(z) + \frac{1}{2^{L-n}} \delta(z-2^{L-n}).
\end{align} 
Here, $n$ is an integer between $0$ and $L$.
The broad distribution in Fig.~\ref{fig:PrZ}(a) is markedly different from this, and is due to the fact that $\rho$ is a non-stabilizer state.

We focus on the non-stabilizer state $\overline{\rho}_{\bs{m}}$ henceforth.
In analogy with random unitary circuits,
we consider the circuit average of $\mathrm{Pr}(z; C_{\bs{m}})$, % probability distribution function
\begin{align}
\label{eq:def_Pr_z}
    \mathrm{Pr}(z) \coloneqq&\ \mathbb{E}_{C_{\bs{m}}} \mathrm{Pr}(z; C_{\bs{m}}) \nn
    =&\ 
    \frac{1}{D} \sum_{x \in \{0,1\}^L}
    \mathbb{E}_{C_{\bs{m}}}
    % \int \mc{D} C_{\bs{m}} \ 
    \delta ( z - \mu(x; C_{\bs{m}}, \rho) \cdot D ) \nn
    =&\ 
    \mathbb{E}_{C_{\bs{m}}}
    % \int \mc{D} C_{\bs{m}} \ 
    \delta ( z - \mu(x; C_{\bs{m}}, \rho) \cdot D ).
\end{align}
Here, after circuit averaging $\mathrm{Pr}(z)$ does not depend on the bitstring $x$ despite the notation, and we can choose $\ket{x} = \ket{0}^{\otimes L}$, for concreteness.
% The circuit-averaged function $\mathrm{Pr}(z)$ is expected to characterize the ensemble $\{\mu (x; C_\bs{m}, \rho) : x \in \{0,1\}^{L} \}$ in a typical instance of the circuit, $C_{\bs{m}}$.

In the unitary limit $p=0$, there are no measurements, and $\mathbb{E}_{C_{\bs{m}}} = \mathbb{E}_{U}$.
Here $\mathrm{Pr}(z)$ should be the Porter-Thomas distribution since the Clifford group forms a unitary 2-design,
\begin{align}
\label{eq:Pr_z_p=0}
    \mathrm{Pr}(z) = \mathbb{E}_{U} \delta ( z - \mu(x; U, \rho) \cdot D ) = e^{-z}.
\end{align}
For $p > 0$, we observe numerically that (see Fig.~\ref{fig:PrZ}(b))
\begin{align}
\label{eq:Pr_z_p>0}
    \mathrm{Pr}(z) \propto \alpha \delta(z) + \beta z^{-\gamma}, \quad \gamma \approx 4.
\end{align}
Since this function $z^{-\gamma}$ diverges as $z \to 0$, the asymptotics is only valid for $z$ greater than some (possibly $L$-dependent, see below) cutoff $\lambda$.
We suspect that the exponent $\gamma$ is universal (as we have checked for a few values of $p$), while the constants of proportionality $\alpha, \beta$ are $\lambda$-dependent (to keep $\mathrm{Pr}(z)$ normalized) and  nonuniversal.

Since the distributions in Fig.~\ref{fig:PrZ}(a,b) have long tails  -- meaning that in a given $C_{\bs{m}}$ the bitstrings occur with rather uneven probabilities -- predicting which ones occur more commonly should be hard, and it is tempting to conjecture the classical hardness of sampling $x$ from the probability distribution $\mu(x; C_{\bs{m}}, \rho)$, for a generic (non-stabilizer) initial state $\rho$.
Given that on a noiseless quantum computer we can simulate the hybrid circuit and produce the state $\rho_{\bs{m}}$, such hybrid circuits may serve the purpose of demonstrating quantum advantage.

However, there is an important caveat here.
% \MF{Why in a footnote??} \YL{I think this is a diversion, being a technical detail.}
% % As it stands, this task also appears hard for a quantum simulation of the hybrid circuit.
As evident from the definition of $\mu$, for a fixed $C$ the bitstring distribution as obtained from measuring $\rho_{\bs{m}}$ still has an explicit dependence on $\bs{m}$.
In each run of the circuit, one gets a new $\bs{m}$, and the bitstring distribution $\mu$ changes from run to run.
Thus, even the circuit itself cannot effiently sample $\mu(x; C_\bs{m}, \rho)$ for any given $\bs{m}$, for we have no control over $\bs{m}$, and cannot repeatedly prepare $\rho_{\bs{m}}$.
To sample $x$ from $\mu(x; C_\bs{m}, \rho)$ for a given $\bs{m}$, it seems that we must again resort to postselection.
% We have not found a way around postselection.

It might be possible to avoid the need of postselection by focusing on a particular subset of non-stabilizer initial states $\rho$, for which the bitstring distributions  $\mu(x; C_\bs{m}, \rho)$ for different $\bs{m}$ can be related to each other by a change of variable in $x$.
Characterizations of such $\rho$ is beyond the scope of this work, which we will discuss elsewhere.

\end{document}